\documentclass[journal]{IEEEtran}

\usepackage{cite}
\usepackage{multicol,lipsum}
\usepackage{amsmath,amssymb,amsfonts}
\usepackage{algorithmic}
\usepackage{graphicx}
\usepackage{textcomp}
\usepackage{xcolor}
\usepackage{booktabs}
 \usepackage{hyperref}
\usepackage{enumerate}
\usepackage{algorithm}
\usepackage{multirow}
\usepackage{standalone}
\usepackage{tikz}
\usetikzlibrary{positioning}
\usetikzlibrary{decorations.markings}
\usepackage{caption}
\usepackage{subcaption}
 \usepackage{makecell}
\usepackage{mathrsfs}  
\ifCLASSINFOpdf
 
\else
 
\fi

\begin{document}
%
\title{ Frame-Based AFDM-ISAC Waveform Design With Chirp-Enabled Pulse Compression }

\author{Qu Luo, ~\IEEEmembership{Member,~IEEE,}
  Zilong Liu, ~\IEEEmembership{Senior Member,~IEEE,}
  Musavian, Leila, ~\IEEEmembership{Member,~IEEE,}
  Thomos, Nikolaos, ~\IEEEmembership{Senior Member,~IEEE,}
  Qihao Peng, ~\IEEEmembership{Member,~IEEE,}
  and
 Pei Xiao, ~\IEEEmembership{Senior Member,~IEEE.}
  
  \thanks{ This work was supported in part by the UK Engineering and Physical Sciences Research Council under Grant EP/X013162/1. The work of Z. Liu, L. Musavian, and N. Thomos was supported in part by the UK Engineering and Physical Sciences Research Council under Grants EP/Y037243/1 (`TITAN/REVOL6G' and `TITAN/LEGEND6G'), EP/X040569/1 (`HASC/RETHIN6G'), EP/X035352/1 (`DRIVE'), EP/Y000986/1 (`SORT'), and EP/X012204/1 (`PerCom'). (Corresponding author: Zilong  Liu)}
  
  \thanks{  Qu  Luo, Qihao Peng and  Pei Xiao  are  with the  5G \& 6G  Innovation Centre, University of Surrey, U. K. (email: \{q.u.luo,  q.peng, p.xiao \}@surey.ac.uk).
 }  
   \thanks{
Zilong   Liu, Musavian, Leila, Thomos, and Nikolaos are   with   the   School   of   Computer   Science   and   Electronics   Engineering,   University   of   Essex,   U. K. (email:  \{zilong.liu,leila.musavian, nthomos\}@essex.ac.uk).}}
\maketitle

\begin{abstract} 
This paper proposes an Affine frequency division multiplexing (AFDM)-empowered integrated sensing and communications (ISAC) design, referred to as AFDM-ISAC.  We first design a novel AFDM-ISAC frame structure that consists of both ISAC and pure data symbols. Each ISAC symbol consists of  a single chirp subcarrier  for both sensing and channel estimation,  while the remaining subcarriers are allocated for communication. Building upon  this structure, we present an analog-domain sensing receiver that down-mixes the received echo with a local chirp to fully exploit \textit{chirp compression} gains avoiding the need  for full-duplex hardware. In addition,  a sensing fusion algorithm, guided by AFDM modulation parameters, is further proposed in the digital domain. 
Leveraging  the distinct features of the proposed AFDM-ISAC frame,  we present  a low-complexity channel estimation scheme for high mobility channels based on a generalized complex exponential basis expansion model (GCE-BEM), along with an optimal  power allocation strategy between pilot and data symbols. Moreover, to support frame-based AFDM communications,   a GCE-BEM-based Kalman filter is also employed  for robust intra-frame channel estimation.
 Simulation results demonstrate the effectiveness and superiority of the proposed schemes in terms of system flexibility, hardware complexity, and ISAC performance.

\end{abstract}


%
\IEEEpeerreviewmaketitle

 \section{Introduction}

\IEEEPARstart{L}{ow}-altitude wireless networks (LAWNs), which interconnect aerial platforms such as unmanned aerial vehicles (UAVs), electric vertical takeoff and landing (eVTOL) aircraft, aerial robots, and other near-ground airborne nodes, have emerged as a critical enabler for future wireless communications \cite{yuan2025ground,AFDMLAE}. By offering flexible deployment, rapid mobility, and three-dimensional connectivity, LAWNs effectively support a wide range of   applications, including emergency communications, traffic surveillance, environmental monitoring, and smart-city sensing. These applications impose a strong demand for both sensing and reliable communication services within LAWNs \cite{11017717}. To meet these requirements, integrated sensing and communication (ISAC) has gained significant attention  by   co-designing    radar and communication functionalities   under  a unifying  wireless system \cite{11017717,sui2025multi,luo2026chirp}. Such an integration leads to improved spectral and energy efficiencies, reduced hardware and storage costs, as well as mutually enhanced sensing and communication functionalities. Among   many research topics, waveforms are at the very heart of ISAC system design,   attracting significant attention from both academia and industry \cite{LiuJSAC,yin2025affine}.

 
Orthogonal frequency-division multiplexing (OFDM) has been studied as  one of the   dominant ISAC waveforms   owing to its prevalence in modern   wireless   systems, such as  long-term evolution, 5G New Radio (NR), and Wi-Fi networks \cite{LiuJSAC}.  However, OFDM may suffer  from significant inter-carrier interference and performance degradation in high mobility channels due to the loss of subcarrier orthogonality.  Motivated by this problem,  orthogonal time frequency space (OTFS) modulation was proposed, through which one can convert a    time-varying multipath channel  into quasi-static representations in the delay-Doppler domain \cite{OTFSWCNC,liu2022near}. Another promising direction is   chirp-based multicarrier   waveforms, such as orthogonal chirp division multiplexing (OCDM) \cite{ouyang2016orthogonal} and affine frequency division multiplexing (AFDM) \cite{BemaniAFDM,RouAFDM,boudjelal2025redefining}. In particular, AFDM  generalizes OFDM and OCDM by introducing a tunable chirp rate, enabling adaptive delay-Doppler resilience and multipath separation in the affine Fourier transform (AFT) domain \cite{zheng2024channel}.  With this flexibility and excellent backwards-compatibility, AFDM offers a favorable trade-off between error rate performance  and implementation efficiency. Importantly, AFDM is able to achieve full channel diversity with reduced pilot overhead and implementation complexity \cite{wang2025afdm}. These advantages make  AFDM as a promising waveform      for ISAC and high-mobility LAWNs   \cite{AFDMLAE,ranasinghe2024joint,luo2024afdm,tao2025affine,li2025affine,wang2025low}.

 \subsection{Related Works}

To date, the research  on AFDM-based ISAC is still in its early stage. There are monostatic and bistatic ISAC systems, whereby the transmitter and radar receiver are co-located and spatially separated, respectively \cite{xia2025power}.
 The authors in  \cite{zhu2024afdm} investigated a bistatic sensing-aided channel estimation scheme and analyzed the ambiguity function (AF) properties  of AFDM. However, their approach cannot estimate the target's velocity. Subsequently,    \cite{ranasinghe2024blind} extended bistatic AFDM sensing to static target scenarios.

Unlike  \cite{zhu2024afdm,ranasinghe2024blind}, most existing AFDM-ISAC studies focused on monostatic sensing architectures. For example,  \cite{ni2022afdm}  introduced a matched-filter-based method combined with a fast cyclic-correlation radar algorithm to estimate target range and velocity in both   time and discrete affine frequency time  (DAFT) domains.
Later,  an improved monostatic AFDM-ISAC scheme that requires only a single AFDM symbol  { was proposed} in \cite{BemaniAFDMisac}.  After performing   self-interference cancellation (SIC) in the DAFT domain, target parameters are estimated via an approximate maximum-likelihood (ML) algorithm.  While the SIC mechanism effectively simplifies hardware implementation, the ML estimation remains computationally intensive due to its reliance on an exhaustive two-dimensional ($2$D) grid search.
 Unlike \cite{ni2022afdm} and \cite{BemaniAFDMisac} focusing   primarily on radar sensing design, \cite{ranasinghe2024joint} proposed a unified framework for joint channel estimation, data detection, and radar parameter estimation over doubly dispersive channels.

More recently, the AF and sensing performance metrics of AFDM-ISAC systems have been extensively   studied in \cite{rou2025normalized,YinAmg,NiAFafdm,ni2025integrated,bedeer2025ambiguity}.   The auto- and cross-AF of   AFDM chirp subcarriers were analyzed   in   \cite{YinAmg}, while   the AF of AFDM under pulse shaped random   signaling was derived in \cite{NiAFafdm}.   In \cite{ni2025integrated}, two novel metrics, i.e., sensing spectral efficiency (SE)   and sensing outage probability, were introduced to characterize the sensing communication trade-offs. Closed-form expressions for the Cramér–Rao lower bounds (CRLBs) and the AF for pilot-assisted AFDM waveforms were further derived in \cite{zhang2025afdm}.   In addition, the authors in  \cite{bedeer2025ambiguity} derived a closed-form expression for the AF of AFDM waveforms modulated with $M$-ary quadrature amplitude modulation (QAM) and identified a chirp-rate condition that minimizes sidelobe levels in the delay/range domain. While the above works primarily focused  on target parameter estimation (e.g., range and velocity),    an AFDM-ISAC framework that estimates the multipath power–delay profile to enhance communication performance was proposed in \cite{xiao2025multipath}. Furthermore, in contrast to prior studies such as \cite{ni2022afdm,BemaniAFDMisac,luo2025target,zhu2024afdm,ranasinghe2024blind,bedeer2025ambiguity,ni2025integrated,xiao2025multipath,zhang2025afdm}, which mainly considered  far-field targets,    a joint angle-delay-Doppler estimation scheme for AFDM-ISAC systems operating in mixed near-field and far-field environments was developed in \cite{luo2025novel}.

\subsection{Motivations and Contributions}

One critical challenge in monostatic ISAC system design is that SIC is often overlooked \cite{BemaniAFDMisac}. Since the transmitter and receiver are co-located, the transmit  waveform may leak into the receive chain, causing severe self-interference and posing a major barrier to practical monostatic radar implementations.   Note that most AFDM- and OTFS-based ISAC designs adopt a communication-centric architecture by largely reusing the communication processing chain \cite{ni2022afdm,ranasinghe2024blind,ni2025integrated,xiao2025multipath,zhang2025afdm,ZegrarISAC,OTFSisac}. Such designs rely on the strong assumption of ideal SIC, which in turn requires costly full-duplex interference cancellation techniques.   Additionally, while chirp signals are well known for their \textit{pulse compression} capability and hence  are widely used in conventional radar systems, this advantage has yet to be fully exploited in AFDM-based ISAC designs, despite some preliminary attempts in \cite{BemaniAFDMisac}.
 Although there are many
works on AFDM in recent years, limited work is done on AFDM frame structure which is a
key step for its practical applications, especially for its applications in ISAC. Furthermore,
conventional channel estimation approaches for doubly selective channels are tedious and
suffer from significant training overhead, thus leading to significant compromise of SE.   Against these challenges, this paper investigates a frame-based AFDM-ISAC scheme in terms of sensing and channel estimation. 

 The  main   contributions of this paper are summarized as follows:

\begin{itemize}
    \item 
     We first propose a novel AFDM-ISAC frame structure composed of ISAC symbols and pure data symbols. In each ISAC symbol, a single  subcarrier is dedicated as the sensing and pilot subcarrier (SPS), while the remaining subcarriers are allocated for communication purpose or used as guard bands. Since the SPS sweeps the entire AFDM bandwidth, it preserves the range resolution equivalent to that of a full AFDM symbol. The received echo of the SPS, together with the interference introduced by the data subcarriers   are then analyzed in detail.

    \item  We further introduce a dedicated analog-domain sensing receiver, equipped with a judiciously designed low-pass filter (LPF). We show that the interference from  the data subcarriers can be well  suppressed,  thereby eliminating the need for a complex and costly full-duplex design. By placing the analog-to-digital converter (ADC) after the LPF, saturation caused by transmitter leakage can be alleviated, since the LPF suppresses high-frequency leakage components before digitization.     In addition,  an AFDM modulation parameter, i.e., chirp slop, oriented   sensing processing algorithm is proposed.    We show that the proposed AFDM-ISAC frame can flexibly balance sensing and communication requirements, such as the sensing performance, the maximum sensing range and speed, communication efficiency, and communication performance.
    \item 
    Building upon the proposed AFDM-ISAC frame structure, a generalized complex exponential basis expansion model (GCE-BEM) is used to approximate the time varying channels, thus achieving accurate channel estimation with  reduced computational complexity.   We derive the signal-to-interference-plus-noise ratio (SINR) of the GCE-BEM-assisted  channel   estimation   to guide the power allocation between the pilot and data subcarriers. Furthermore,  to address the channel variations of  the proposed AFDM-ISAC signaling,   a  Kalman filter based
exponential basis expansion model (KF-BEM)    is proposed for efficient channel tracking. 
    \item We conduct  extensive simulation results to demonstrate the sensing and communication superiority of the proposed AFDM-ISAC.  We show that the proposed AFDM-ISAC solutions outperform the conventional  OTFS-, OFDM- and AFDM-based schemes in terms of the range and speed root mean square error (RMSE). It is shown that selecting AFDM parameter $c_1 \in \left\{ \frac{2}{2N}, \frac{3}{2N}, \frac{4}{2N} \right\}$, along with a pure-data-to-ISAC symbol ratio $\eta \in \{1, 2, 3 \}$, yields a   desirable trade-off between sensing accuracy and communication performance.
\end{itemize}

 The remainder of this paper is  outlined as follows.
Section II introduces the basic  AFDM principles. The proposed AFDM-ISAC frame structure design  
is introduced  in Section III.     Section IV and  Section V present  the proposed sensing   and communication receiver design, respectively. Section VI evaluates the sensing and communication performance of the proposed AFDM-ISAC, followed by the conclusions in Section VII. 
 
\subsection{Notation}

  $\mathbb{C}^{k\times n}$ denotes  the $(k\times n)$-dimensional complex  matrix. ${{\mathbf{I}}_{n}}$ denotes an $n \times n $-dimensional  identity matrix.   $\text{diag}(\mathbf{x})$ gives a diagonal matrix with the diagonal vector of $\mathbf{x}$. $(\cdot)^\mathcal T$, $(\cdot)^ \dag $ and $(\cdot)^\mathcal H$ denote the transpose, the conjugate and the Hermitian transpose operation, respectively.  $\|\mathbf{x}\|_2$ and $|x|$ return the Euclidean norm of vector $\mathbf{x}$ and the absolute value of $x$, respectively. $\mathbb{C}$ and $\mathbb Z$ denote the complex and integer spaces, respectively.  $ \left < \cdot \right>_N$ denotes the modulo-$N$ operation,  and   $\mathcal{CN}(0,1) $ denotes the complex distribution with zero mean and unit variance.

\begin{figure*}
    \centering
\includegraphics[width=1\linewidth]{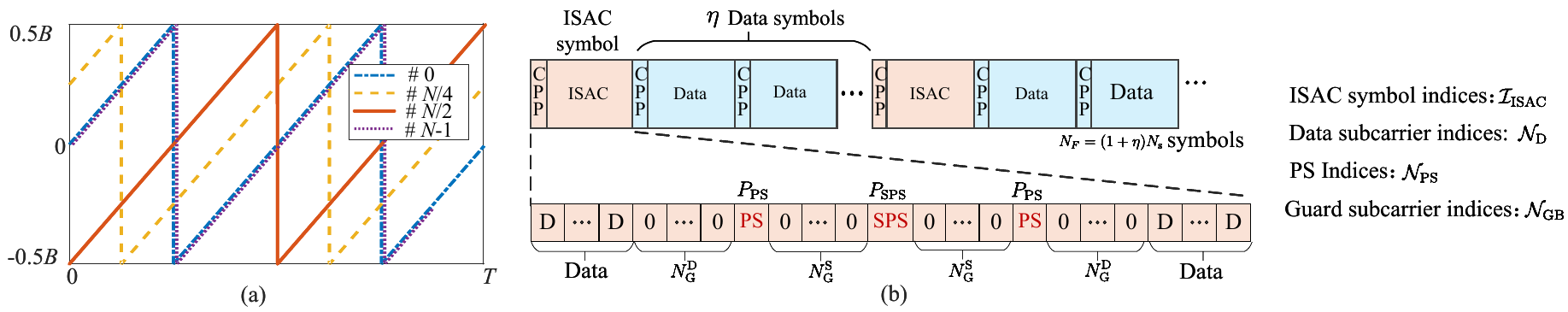}
    \caption{Illustration of (a)  wrapped AFDM subcarriers ($K=2$) and (b) the   proposed AFDM-ISAC frame structure.}
    \vspace{-1.5em}
    \label{AFDMTF}
\end{figure*}

\section{Introduction to Basic Principles of AFDM}
This section introduces the time–frequency representation and provides a detailed mathematical formulation of AFDM subcarriers, which serves as the foundation for the proposed AFDM-ISAC design. The AFDM subcarriers are constructed based on the AFT, whose kernel is defined as  \cite{BemaniAFDM}
 \begin{equation}
 \small
 K_{a,b,c,d}(t, u) = \frac {1}{\sqrt {2\pi \vert b\vert }}{{\rm e}^{-j \left({{\frac{a}{ 2b}}u^{2}+{\frac{1}{ b}}ut+{\frac{d}{ 2b}}t^{2}}\right)}},
 \end{equation}
where  $(a,b,c,d)$ are the AFT parameters and 
$t$ and $u$ denote the time- and affine-domain variables, respectively.  By sampling the transform kernel  at the AFT domain with a   sample interval  $ \Delta u$, one has  
 \begin{equation}
 \small
 \label{phi1}
 K_{a,b,c,d}(t, u) \vert_{u = n \Delta u} = \frac {1}{\sqrt {2\pi \vert b\vert }}{{\rm e}^{j\left({{\frac{a}{ 2b}}n^2 \Delta u^{2}+{\frac{1}{b}}n \Delta u t+{\frac{d}{ 2b}}t^{2}}\right)}}. 
 \end{equation}
Let $\Delta t =T/N$ denote the sampling interval in the time domain, where $T$ is the signal duration and $N$ the total number of samples. Following \cite{BemaniAFDM}, the AFDM modulation parameters are defined as $c_{1} = \frac {d}{4\pi b}\Delta t^{2}$ and $c_{2} = \frac {a}{4\pi b}\Delta u^{2}$. The transform is reversible if $\Delta t \Delta u = \frac {2\pi | b|}{N}$ holds \cite{BemaniAFDM}.    The AFDM defined in \cite{BemaniAFDM} uses the first type as the transform kernel. To align with the signal model in \cite{BemaniAFDM}, we consider  $b>0$  as well.   Under the constraints of $\Delta t \Delta u = \frac {2\pi | b|}{N}$  and $b>0$,    (\ref{phi1}) reduces to 
 \begin{equation}
 \small
 \label{kn}
 \begin{aligned}
    K_{n}(t)&  = {{\rm e}^{j 2 \pi  \left({ {\frac{c_1 N^2}{ T^2}}t^{2}   +\frac{   n  t}{T} + c_2n^2   } \right)}},  b>0,
 \end{aligned}
 \end{equation}
which forms a set of $N$ orthogonal chirp signals, i.e., $\{ K_{n} (t) \}^{N-1}_{n=0}$.

After sampling $K_{n}(t)$ with period $\Delta t$, the AFT kernel in the digital domain can be expressed as
\begin{equation}
\small
\Psi_n(m)  ={1\over \sqrt{N}}e^{j2\pi \left ( c_{1}m^{2}+c_{2}n^{2}+{nm\over N} \right)}.
\label{pjidig}
\end{equation}
By collecting $N$ normalized samples
of each signal in (\ref{pjidig}), we
can obtain the so-called DAFT matrix. Namely,  $\mathbf A = \Lambda_{c_2}  \mathbf F \Lambda_{c_1}  $ with  $\Lambda_{c} =  \text{diag}\left ( e^{-j2\pi cn^2}, n=0,1, \ldots, N-1 \right)$, where    $\mathbf F$ denotes the $N \times N$ DFT  matrix with entries  $[\mathbf{F}]_{m,n} = e^{-j2\pi mn/N} / \sqrt{N}$.

\textit{Remark 1: Let $B= \frac{1}{\Delta t}= \frac{N}{T}$. It should be noted that $K_{n}(t), 0 \leq n \leq N-1$, occupies a bandwidth of $2c_1NB$ within the time duration $T$. However, according to the sampling theorem, time-domain sampling introduces spectral wrapping in the digital AFDM signal, thereby limiting the effective bandwidth to the range $[-0.5B, 0.5B]$. Fig.~\ref{AFDMTF}(a) illustrates an example of the  wrapped time–frequency representation with  $c_1=\frac{1}{N}$.}

In the sequel, we derive the detailed expressions of the time-domain wrapped AFDM subcarriers.  Define $\phi_n(t) = \frac{c_1 N^2}{ T^2}t^{2} +\frac{ nt}{N}B$, whose derivative with respect to $t$ is $\phi_n^{'}(t) = \frac{2c_1 N^2}{ T^2}t +\frac{ n}{N}B$. 
Note that  $\phi_n(t)$ describes the phase changes of the $ K_{n}(t)$  with $\phi_n^{'}(t)$  being the changing rate.  Then, the wrapped AFDM subcarriers can  be expressed as
  \begin{equation}
  \small
  \label{eqPhi}
 \begin{aligned}
 \Psi_n(t)  =    {{\rm e}^{ j 2 \pi  \left(  { \int_{0}^{t} \bar{{\phi}}_n^{'}(x)dx    + c_2n^2  }\right)  }},
 \end{aligned}
 \end{equation}
 where $ \bar{{\phi}}_n^{'}(x) =  \left < \phi_n^{'}(x)+ \frac{B}{2}\right>_B-\frac{B}{2}$.   Later, we will show that by judiciously aggregating  the wrapped AFDM subcarriers,  it enables  enhanced range and velocity estimation.


\begin{figure*}
    \centering
    \includegraphics[width=1\linewidth]{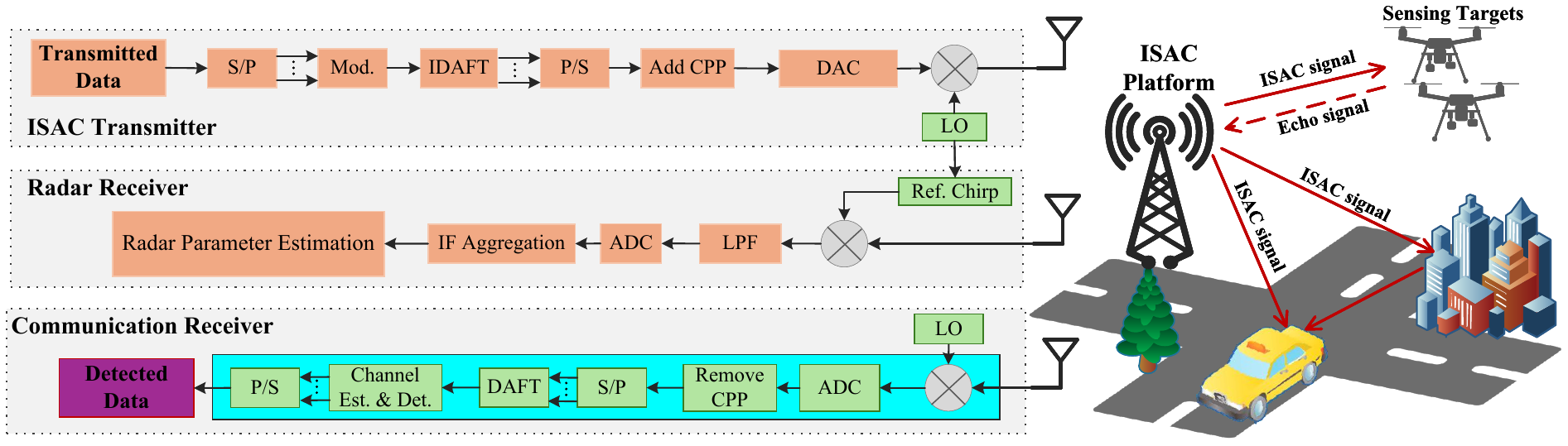}
    \caption{The proposed AFDM-ISAC signal processing.}
\vspace{-1.5em}
    \label{PropAFDMisac}
\end{figure*}

\textit{Remark 2: The OCDM system \cite{ouyang2016orthogonal} defines time-domain chirp subcarriers as $K_{n}(t) = e^{-j\pi/4}{\rm e}^{-j \pi {\frac{ N}{ T^2}} \left({ t -\frac{ T}{N} n }\right)^2}$.  In fact, for $c_1=c_2=-1/2N$ and $b>0$, (\ref{kn}) simplifies to      
  \begin{equation}
  \small
 \begin{aligned}
           K_{n}(t)  =  {{\rm e}^{-j 2 \pi  \left({  \frac{n^2}{2N}   -\frac{  n  t}{T}+{\frac{ N}{2 T^2}}t^{2}}\right)}} 
     = {{\rm e}^{-j \pi  {\frac{ N}{  T^2}} \left({  t   -\frac{     T}{N} n }\right)^2}},
 \end{aligned}
 \end{equation}
which is essentially equivalent to the OCDM subcarriers in \cite{ouyang2016orthogonal}, except for a fixed phase shift $e^{-j\pi/4}$.
}

 \section{The Proposed AFDM-ISAC Signal Structure}

In this section, we first introduce the proposed AFDM-ISAC frame structure. We then proceed to describe the    sensing and communication signal models.

\subsection{The Proposed AFDM-ISAC Signal Model }

The proposed AFDM-ISAC frame structure is illustrated in Fig.~\ref{AFDMTF}(b). It consists of ISAC symbols and pure data symbols. An ISAC symbol is employed for both sensing and communication, whereas a data symbol is dedicated solely   to communication. Each frame contains $N_{\text{s}}$ ISAC symbols, which are periodically placed at positions of
\begin{equation}
 \small
   \mathcal I_{\text{ISAC}} = \{0, \eta+1, \ldots, (N_{\text{s}}-1)(\eta+1)\},
\end{equation} 
where $\eta \in \mathbb Z$  denotes the ratio of pure-data  symbols to  ISAC symbols. Therefore, $\eta$ controls the trade-off between the number of ISAC and data symbols. In particular, when $\eta = 0$, the AFDM-ISAC frame only consists  of ISAC symbols. Accordingly, the total number of AFDM symbols is given by $N_F = (1+\eta)N_{\text{s}}$.

As shown in Fig.~\ref{AFDMTF}(b), each ISAC symbol contains sensing and channel estimation subcarriers, communication subcarriers, and guard subcarriers.    The $(N/2)$th  subcarrier, $\Psi_{\frac{N}{2}}^{k}(t)$, is designated as the SPS.  The reasons are twofold: 1) Since each AFDM subcarrier spans the entire bandwidth, using a single chirp subcarrier for sensing can still preserve the sensing resolution. Later, we will show that the proposed design also leads to a low-complexity sensing receiver and improved communication SE; 2) According to Remark 1, the $(N/2)$th AFDM subcarrier starts from $-0.5B$ and sweeps the full bandwidth $K$ times for $c_1 = \frac{K}{2N}$ with $K \in \mathbb{Z}$. For dechirping-based sensing receiver design, continuous chirp signals are desired. 
To further improve channel estimation accuracy, two additional pilot subcarriers (PS) are also inserted. Although a single subcarrier can capture the entire channel in AFDM, it relies on a large $c_1$ value to ensure that multipaths with fractional Doppler are well separated in the DAFT domain for accurate estimation \cite{MIMOAFDMchannest}. Hence, we adopt a multiple-pilot structure to mitigate the interference caused by fractional Doppler across different paths. It is also worth noting that the communication pilot guard is much shorter than the sensing guard. To further reduce the number of channel parameters to be estimated, the BEM model will be introduced  in Section VI. In addition, guard-band subcarriers are placed to suppress interference. Let $N_{\text{G}}^{\text{S}}$ and $N_{\text{G}}^{\text{D}}$ denote the numbers of guard subcarriers between the SPS and PS, and between the PS and data subcarriers, respectively. The number of communication subcarriers in each ISAC symbol is then given by
 \begin{equation}
  \small
      N_{\text{C}} =  N- 2N_{\text{G}}^{\text{D}}-2N_{\text{G}}^{\text{S}}-3.
 \end{equation}
The modulated data of the $i$th   symbol,  $ i \in \mathcal I_{\text{ISAC}}$,  in the DAFT domain is given by
\begin{equation}
\small
\label{xn}
 x_{i,n} =   \left\{\begin{matrix}
 \sqrt{P_{\text{SPS}}},&   n= \frac{N}{2},\\
  \sqrt{P_{\text{PS}}},&   n \in \mathcal N_{\text{PS}}\\
 0, &    n \in \mathcal N_{\text{GB}},  \\   
 \text{Data Symbols,} &   n \in \mathcal N_{\text{D}},\\
\end{matrix}\right. 
\end{equation}
where $P_{\text{SPS}}$  and $P_{\text{PS}}$ are the allocated power levels for the  SPS and PS, respectively, and  $\mathcal N_{\text{PS}}=  \{ \frac{N}{2}-N_{\text{G}}^{\text{D}}-N_{\text{G}}^{\text{S}}-1,  \frac{N}{2}+N_{\text{G}}^{\text{D}}+N_{\text{G}}^{\text{S}}+1\} $,  $ \mathcal N_{\text{GB}}$ and $ \mathcal N_{\text{D}}$  are the indices of the PS, guard band, and data subcarriers, respectively.
Accordingly, the continuous-time waveform of the $i$th symbol can be expressed as\begin{equation}
\small
\label{withoutCPP}
 s_{i}(t) =   \sum _{n=0}^{N-1}   x_{i,n}\Psi_n (t-iT). 
\end{equation}
This waveform is transmitted   after adding cyclic prefix (CP).   For digital transmission, the time-domain signal in the $i$th symbol is expressed as $\mathbf s_i = \mathbf A^{\mathcal H} \mathbf x_i \in \mathbb C^{N \times 1}$, where $\mathbf x_i = [x_{i,0}, x_{i,1}, \ldots, x_{i,N-1}]^{\mathcal T}$. Owing to the distinct signal periodicity, a chirp-periodic prefix (CPP) is  employed in the digital domain, i.e., \cite{zhu2023design}
  \begin{equation}
  \small
    s_i[n] =   s_i[N+n] e^{-j 2 \pi c_1(N^2+2Nn)}, n=-N_{\text{CPP}}, \cdots,-1,
\end{equation}
where $N_{\text{CPP}}$ denotes the CPP length. For $c_1 = \tfrac{K}{2N}$ with $K \in \mathbb Z$, the CPP reduces to the conventional CP.

  



\subsection{Sensing  Signal   Model}
 
Let $L$ be the number of sensing targets, with $R_l$ and $v_l$ denoting the range and velocity of the $l$th target, respectively. The received echo signal of the $l$th target has the same form as $s_i(t)$ but with a delay $\tau_l$, defined as \footnote{It is worth noting that this delay model effectively characterizes both delay and Doppler effects.}
 \begin{equation}
 \small
 \label{tau}
     \tau_l  = {2(R_l+v_lt)}/{\nu},
 \end{equation}
where $\nu$ denotes the speed of light. Similar to existing works, we assume line-of-sight propagation \cite{ni2022afdm,BemaniAFDMisac,luo2025target,zhu2024afdm,ranasinghe2024blind,bedeer2025ambiguity,ni2025integrated,xiao2025multipath,zhang2025afdm}, which is also the predominant channel model for aerial UAV scenarios.  Accordingly, the received echo can be expressed as  

\begin{equation}
\small
\begin{aligned}
\label{rsym}
      r_i(t)  = &   \sum_{l=0}^{L-1}      h_{l}^{\text{S}}  s_i( t-\tau_l)e^{j2\pi f_c( t-\tau_l)}          +w_i(t),
\end{aligned}
\end{equation}
\color{black}
where $h_{l} $ denotes the channel gain of the $l$th target, $w_i(t) \sim \mathcal{CN}(0,\sigma_s)$, and $f_c$ is the carrier frequency. Note that $h_{l} $ accounts for both channel attenuation and radar cross-section attenuation \cite{xia2025power}.

\subsection{Communication Signal  Model}
We consider a doubly selective channel between the ISAC platform and the communication user. For simplicity, its equivalent discrete channel impulse response is adopted, with the expression at time index $n$ and tap $p$ of the $i$th AFDM symbol given by
 \begin{equation}
 \small
 \label{DFIR}
     h_i(n,p)=\sum _{p=0}^{P-1}  {h}_{i,p}e^{-j 2\pi f_{i,p}n}\delta (p - p_{i}),
 \end{equation}  
 where $P$ is the number of paths, $ {h}_{i,p}$ and $f_{i,p}$ denote the channel fading coefficient  and Doppler shift of the $p$th path in the $i$th AFDM symbol, respectively.      Due to user mobility, the physical channel taps are time-varying.  By assuming a  channel taps remain unchanged within a given AFDM symbol but vary across different symbols. Let $\mathbf h_i = [{h}_{i,1},{h}_{i,2},\ldots,{h}_{i,P} ]^{\mathcal T}$.   According to the wide-sense stationary uncorrelated scattering  model, the time-varying $\mathbf h_i$ can be approximated by a first-order auto regression (AR) model \cite{komninakis2002multi,4244643}.
\begin{equation}
\small
\label{hsv}
    \mathbf h_{i+1} =  \widetilde{ \mathbf S} \mathbf h_{i} + \widetilde{ \mathbf v} _i,
\end{equation}
where $\widetilde{ \mathbf S}  \in \mathbb C^{P \times P}$ denotes the state transition matrix which is  at the receiver \cite{4244643}, and $\widetilde{ \mathbf v} _i \in   \mathcal {CN}(\boldsymbol 0, \sigma_v \mathbf I_P)$.

 Note that (\ref{DFIR}) can be expressed in matrix form as
\begin{equation}
\small
\label{Hi}
\setlength{\arraycolsep}{0.8pt}
    \begin{aligned}
        {{\mathbf{H}}_{i}}\! =\! \left [{\!{\begin{array}{cccccc} {h_{i}(0,0)}&0&\cdots &{h_{i}(0,P\!-\!1)}&\cdots &{h_{i}(0,1)}\\ {h_{i}(1,1)}&{h_{i}(1,0)}&0&\cdots &\cdots &{h_{i}(1,2)}\\ \vdots &\ddots &\ddots &\ddots &\ddots &\vdots \\ 0&\cdots &0&{h_{i}(N\!\!-\!\!1,P\!\!-\!\!1)}&\cdots &{h_{i}(N\!\!-\!\!1,0)} \end{array}} \!}\right].
    \end{aligned}
\end{equation}
  Then, the received signal  of the $i$th AFDM symbol in the discrete AFT domain is given by
\begin{equation}
\small
\label{yi_1}
\begin{aligned}
        \mathbf y_i & =  \mathbf A \mathbf {H}_{ i}\mathbf A^{\mathcal H}\mathbf {x}_i + \mathbf {n}_i \triangleq   \mathbf {H}_{\text{eff},i} \mathbf {x}_i + \mathbf {n}_i, 
\end{aligned}
\end{equation}
where $\mathbf n_i \in \mathcal {CN}(0, \sigma_c \mathbf I_N)$ denotes the Gaussian noise and $\mathbf {H}_{\text{eff},i} = \mathbf A \mathbf {H}_{ i}\mathbf A^{\mathcal H} $ denotes the effective channel matrix.


\section{The Proposed Sensing Receiver Design}

Building upon the aforementioned  AFDM-ISAC signal model, this section presents a novel  sensing receiver, including the analog-domain sensing process and the  digital-domain parameter  estimations, as shown in Fig. \ref{PropAFDMisac}. Finally, the detailed performance indicators are presented.  Since only the ISAC symbols are relevant to the sensing receiver, we reuse the subscript $i$ in this section to denote the index of the $i$th ISAC symbol, i.e., $i \in \{0, 1, \ldots, N_{\text{s}}-1\}$.
 
\subsection{Sensing Process}

We denote $t_s$ as the start time of the $i$th ISAC symbol and re-define $t$ as follows:
\begin{equation}
\small
\label{tSym}
    t = i T_{\text{s}}+t_s,   \quad 0\leq t_s<T, \quad 0 \leq i \leq N_s -1,
\end{equation}
where $T_{\text{s}} = (1+\eta)(T + T_{\text{CPP}})$, \color{black} with $T_{\text{CPP}} \triangleq \tfrac{TN_{\text{CPP}}}{N}$ denoting the CPP duration. By substituting (\ref{tSym}) into (\ref{tau}), we obtain
 \begin{equation}
 \small
 \label{tau2}
     \tau_l  =    \frac{2(R_l+v_l(iT_{\text{s}}+t_s))}{\nu}.
 \end{equation}
Accordingly, based on (\ref{rsym}), the received echoes of the $N_F$ symbols can be expressed as

\begin{equation}
\small
\begin{aligned}
      r(t)  
         =  \sum_{i=0}^{N_{\text{F}} -1}   \sum_{l=0}^{L-1}      h_{l}^{\text{S}}  s_i( t-\tau_l)e^{j2\pi f_c( t-\tau_l)}          +w(t).
\end{aligned}
\end{equation}
\color{black}

 On the other hand, the chirp synthesizer of the radar receiver, which shares the same local oscillator (LO) with the AFDM-ISAC transmitter, generates  the conjugated version of the SPS   as the reference chirp.  A delay is introduced to this local reference signal for sensing purpose. Specifically, the $i$th reference signal generated by the chirp synthesizer from the $N/2$-th conjugated AFDM SPS is expressed as
\begin{equation}
\begin{aligned}
\small
\label{reflochirp}
    &  s_{\text{ref},i} (t_s-\tau_{\text{ref}}) = \Psi^{*}_{\frac{N}{2}}\left( t_s-\tau_{\text{ref}}\right)e^{-j2\pi f_c  (t_s-\tau_{\text{ref}})} ,
\end{aligned}
\end{equation}
where $\tau_{\text{ref}} = \frac{2R_{\text{ref}}}{\nu}$, with $R_{\text{ref}}$ denoting the sensing range of interest. The parameter $R_{\text{ref}}$ will be discussed in more detail later. Upon receiving the echo signal, $r(t)$ is mixed with $s_{\text{ref},i}(t)$ to generate the intermediate-frequency (IF) signal, i.e.,
    \begin{equation}
     \small
    \label{Mix1}
    \begin{aligned}
       & r_i^{\text{IF},l}(t) \vert _{t = iT_{\text{s}}+t_s} = r_i(t) s_{\text{ref},i}  (t-\tau_{\text{ref}})  \\ 
      &    \simeq \!    \underbrace{    \sum_{l=1 }^{L} \!   \! \sqrt{P_{\text{SPS}}} h_l^{\text{S}}  \Psi_{\frac{N}{2}} (t \! -\! \tau_l \!- \!\!iT_{\text{s}})  \Psi_{ N/2}^{* } (t\!- \!\tau_{\text{ref}}\!-\!iT_{\text{s}})   e^{-j2\pi f_c ( \tau_l \!- \!\tau_{\text{ref}}) } }_{\text{Desired IF signal}}               \\
&    + Z^{\text{IF}}(t),
    \end{aligned}
\end{equation}
where 
\begin{equation}
    \small
    \begin{aligned}
         Z^{\text{IF}}(t) & = \sum_{l=1}^{L}     \sum_{n \in \{\mathcal N \backslash \frac{N}{2} \}}     h_l^{\text{S}} x_{i,n}\Psi_n  (t-\tau_l-iT_{\text{s}})   \\
         & \quad \quad \Psi_{ N/2}^{* } (t-iT_{\text{s}})   e^{-j2\pi f_c(\tau_l-\tau_{\text{ref}})}    + \widetilde w(t),
    \end{aligned}
\end{equation}
\color{black}
denotes the interference from other chirp subcarriers, $\widetilde w(t) =   s_{\text{ref},i} (t_s-\tau_{\text{ref}})w(t)$ and $ \mathcal N \backslash \frac{N}{2}$ denotes the set of  $N$ subcarriers by excluding the SPS.   As mentioned earlier, the SPS sweeps the bandwidth $K$ times. Based on (\ref{eqPhi}), the $k$th sweeping component of the SPS, denoted by $\Psi_{\frac{N}{2}}^{k}(t)$, can be expressed as
 \begin{equation}
     \small 
     \label{phi_nk}
     \Psi_{\frac{N}{2}}^{k}(t) =  e^{j2\pi \left(\frac{\alpha}{2}t^2 - \left (\frac{1}{2}+k \right)Bt + \frac{N^2}{4}c_2\right)}, k \in \{0,1,\ldots,K\},
 \end{equation}
where      $\alpha \triangleq \frac{c_1 B}{2T}$ is  the chirp rate. 

Next, we focus on analyzing the desired IF signal. In the case of $\tau_l>\tau_{\text{ref}}, \forall l$, Fig.~\ref{SPSecho} illustrates an example of the time–frequency representation of the transmitted SPS and the received echo with $c_1 = \frac{1}{N}$. We categorize the symbol duration into three types of time windows, i.e.,
\begin{equation}
\small
 \left\{\begin{matrix}
 \mathcal T_k^{\text{I}} \! = \! \left\{t_s \vert t_s \in \left  [\frac{kT_{\text{s}}}{K} + \tau, \frac{(k+1)T}{K} + \tau_{\text{ref}}  \right  ]   \right\}, \!\! & \!\!k= 0,  \cdots, K\!\!-\!\!1  \\
 \mathcal T_k^{\text{II}}\!= \!  \left\{t_s \vert t_s \!\in \!\left  (  \frac{(k+1)T }{K} \!\!+\!\! \tau_{\text{ref}}, \frac{(k+1)T }{K} \!\! +\!\! \tau  \right  ]   \right\},\! &\!\! k= 0,   \cdots, K\!\!-\!\!2  \\
 \mathcal T_k^{\text{III}}=  \left\{ t_s  \vert \text{otherwise}\right\}.  & 
\end{matrix}\right. 
\end{equation}

By substituting (\ref{phi_nk}) into (\ref{Mix1}), the IF signal from mixing the $k$th SPS echo component of the $l$th target with the reference $s_{\text{ref},i}(t)$ is given by
       \begin{equation}
\label{mix2}
\begin{aligned}
      & \Psi_{N/2}^k  (t_s -\tau_l)e^{j2\pi f_c(t_s -\tau_l)}     s_{\text{ref},i} (t-\tau_{\text{ref}})\vert_{t_s \in  \mathcal T_c^{\text{I}}}   \\ & =   \Psi^c_{N/2} (t_s -\tau_l)    \Psi_{ N/2}^{k,* } (t_s-\tau_{\text{ref}})   e^{-2\pi f_c\left ( \tau_l-\tau_{\text{ref}}\right) }  \\
          & =   e^{-j2\pi \left( -\frac{\alpha}{2}  \left(\tau_l^2-\tau_{\text{ref}}^2\right) + \alpha t_s \left ( \tau_l-\tau_{\text{ref}}\right) + \left( f_c- \left(\frac{1}{2} +k\right)B \right) \left ( \tau_l-\tau_{\text{ref}}\right)   \right)}  \\
      &   \overset{(\text{i})}{\simeq}   e^{-j2\pi \left( \alpha t_s \left ( \tau_l-\tau_{\text{ref}}\right) + \left( f_c- \left(\frac{1}{2} +k\right)B \right)\left ( \tau_l-\tau_{\text{ref}}\right)  \right)}  \\
        &   \overset{(\text{ii})}{\simeq}   \underbrace{    e^{ -j2\pi \left(  \frac{2\alpha  \Delta R_l}{\nu} 
 + \frac{2v_l}{\nu} \left (f_c- \left(\frac{1}{2} +k\right)B + \alpha iT_{\text{s}}  \right)     \right ) t_s}}_{\text{Beat frequency}} \\
 &   \quad  \underbrace{e^{\frac{-j4\pi v_l \left(f_c- \left(\frac{1}{2} +k\right)B \right)}{\nu}iT_{\text{s}} } }_{\text{Phase change over each symbol}} \underbrace{e^{\frac{-j4\pi \Delta R_l\left(f_c- \left(\frac{1}{2} +k\right)B\right)}{\nu} + \frac{-j4\pi\alpha v_lt_s^2}{\nu}}}_{\text{Phase}},\\
\end{aligned}
\end{equation}
where $t_s \in \mathcal T_k^{\text{I}}$. Step (i) holds due to the fact that  $-\frac{\alpha}{2}(\tau_l^2-\tau_{\text{ref}}^2)$ is relatively small and can therefore be neglected. Step (ii) follows by substituting $\tau$ with (\ref{tau}), where $\Delta R_l = R_l - R_{\text{ref}}$\footnote{The beat frequency (BF) in (\ref{mix2}) refers to the frequency component of the IF signal that carries the target range information. By estimating the BF, the target range can be determined.}. In addition, in (\ref{mix2}), the term $\frac{-j4\pi\alpha v_lt_s^2}{\nu}$ is also small and can be neglected.
Hence, the final phase of the desired IF signal at the $k$th component can be expressed as 
\begin{equation}
\small
\label{IFafdm}
\begin{aligned}
       \text{IF}_{\frac{N}{2},i}^{k} (t_s)  & =  \phi_{0}^{k}   +  f_d^k iT_{\text{s}}  +     \text{BF}^k_{\frac{N}{2},i}t_s, t_s \in  \mathcal T_k^{\text{I}},    \\
 f_d^k & =  \frac{ 2v_l \left (f_c- \left(\frac{1}{2} +k\right)B\right)}{\nu},\\
 \text{BF}^k_{\frac{N}{2},i} &  =    \frac{2\alpha \Delta R_l}{\nu} 
 + \frac{2v_l}{\nu} \left (f_c- \left(\frac{1}{2} +k\right)B + \alpha iT_{\text{s}} \right),   \\
 \phi_{0}^{k} & = \frac{2 \Delta R_l \left(f_c- \left(\frac{1}{2} +k\right)B\right)}{\nu},
\end{aligned}
\end{equation}
where $f_d^k$ and $\text{BF}^k_{\frac{N}{2},i}$ denote the Doppler and BF terms, respectively, and $\phi_{0}^{k}$ represents the initial phase obtained by neglecting the term ${-j4\pi\alpha v_lt_s^2}/{\nu}$ due to its insignificance. Similarly, a mixed signal for the wrapped components, i.e., $t_s \in \mathcal T_c^{\text{II}}$, can be readily obtained. Specifically, the IF signal has the same form as (\ref{IFafdm}) but with different IF values and initial phases  respectively given by
\begin{equation}
\small
\begin{aligned}
    & \overline{ \text{BF}}_{\frac{N}{2},i}^k =   { \text{BF}}_{\frac{N}{2},i}^k -B, \\
    & \overline{\phi}_{0}^{k}  = {2 \Delta R_l \Big(f_c- \Big(\frac{3}{2} +k\Big)B\Big)}/{\nu}.
\end{aligned}
\end{equation}

Note that (\ref{IFafdm}) is obtained by choosing $t_s$ as the common reference start time for all $K$ chirp segments within each ISAC symbol. Since each chirp segment physically starts at a different time instant, we define $t_s^k$ as the local start time of the $k$th chirp segment, i.e.,
\begin{equation}
\small
\label{localt}
    t_s^k \triangleq t_s - \frac{kT}{K},  \quad  0 \leq t_s^k < \frac{T}{K},
\end{equation}
which measures the elapsed time from the beginning of the $k$th segment. Substituting (\ref{localt}) into (\ref{IFafdm}) and utilizing the identity $\alpha  \frac{T}{K} = B$, the IF signal in (\ref{IFafdm}) can be equivalently rewritten as
\begin{equation}
\small
\label{IFafdm2}
\begin{aligned}
       \text{IF}_{\frac{N}{2},i}^{k} (t_s^k)    =  {\phi}_{0}^0   +  {f}_d^0   iT_{\text{s}}  +     {\text{BF}}_{\frac{N}{2},i}^0  t_s^k.
\end{aligned}
\end{equation}
  As shown in (\ref{IFafdm2}), by adopting the  time reference $t_s^k$,   the Doppler, BF, and initial phase are inherently independent of the chirp-sweep index $k$. This property motivates the proposed AFDM-oriented parameter estimation presented in Section~\ref{Secest}.

\color{black}

 \vspace{-1.5em}
 
\subsection{LPF Design}

It should be noted that the noise term  $Z^{\text{LF}}(t)$ in (\ref{Mix1}) may not be Gaussian; instead, it has a similar form to that of (\ref{mix2}). Specifically, $\Psi_n(t-\tau_l-iT_{\text{s}})\Psi_{N/2}^{}(t-iT_{\text{s}}),  n \in \mathcal N_{\text{D}} \cup N_{\text{PS}}$, exhibits an IF signal similar to (\ref{mix2}) but with a different  BF.  This indicates that $Z^{\text{LF}}(t)$ can be effectively suppressed by properly designing a LPF and selecting a suitable guard band. In the following, we present the BF of the PS chirp subcarrier. The BF of the left PS mixed with the SPS, i.e., the BF of $\Psi_{\frac{N}{2}-N_{\text{G}}^{\text{S}}-1}(t-\tau_{\text{ref}}-\tau_l-iT_{\text{s}}) s_{\text{ref},m}^{}(t-\tau_{\text{ref}}-iT_{\text{s}})$, can be expressed as
\begin{equation}
\small
\label{bf1}
 \text {BF}_{\frac{N}{2} \! - \!N_{\text{G}}^{\text{S}}-1} \in  
 \Big \{\frac{2\alpha \Delta R_l}{\nu} \! -\! N_{\text{G}}^{\text{S}}\Delta f,   \frac{2\alpha \Delta R_l}{\nu}-N_{\text{G}}^{\text{S}}\Delta f + B\Big \}.
\end{equation}
Similarly, the BF of the right PS mixed with the SPS can be expressed as
\begin{equation}
\small
\label{bf2}
 \text {BF}_{\frac{N}{2}+N_{\text{G}}^{\text{S}}+1} \in  
 \Big \{\frac{2\alpha \Delta R_l}{\nu} \! + \! N_{\text{G}}^{\text{S}}\Delta f,   \frac{2\alpha \Delta R_l}{\nu} \! + \! N_{\text{G}}^{\text{S}}\Delta f- B \Big \}.
\end{equation}
In addition, since the PSs are the closest  nonzero chirp subcarriers to the SPS, it can be readily verified that the BF of the $n$th communication subcarrier is lower bounded by
 \begin{equation}
    \small
    \label{lpf}
    \text {BF}_{n} \geq \min\left\{ \text {BF}_{\frac{N}{2}-N_{\text{G}}^{\text{S}}-1 }, \text {BF}_{\frac{N}{2}+N_{\text{G}}^{\text{S}}+1}  \right\}, \forall n \in \mathcal N \backslash \frac{N}{2}. 
\end{equation}
 The LPF should be designed to pass the desired IF signal while suppressing unnecessary high-frequency interference as well as the IF signals from communication subcarriers. Denote $f_\text{stop}$ and $f_\text{pass}$ as the stopband and passband of the LPF, respectively. Hence, we have
\begin{equation}
\small
\label{LPfdesign}
\begin{aligned}
    f_\text{pass} & \geq \text {BF}_{\frac{N}{2}, \text{max}}   \equiv \frac{2\alpha \vert \Delta R_{\text{max}}\vert}{\nu} ,  \\
  f_\text{stop} &  \leq   N_{\text{G}}^{\text{S}}\Delta f-\frac{2\alpha \vert\Delta R_{\text{max}}\vert}{\nu}, 
\end{aligned}
\end{equation}
where $\text{BF}_{\frac{N}{2}, \text{max}}$ denotes the maximum BF of the desired IF signal, and $\Delta R_{\text{max}} = \underset{l}{\text{max}}(R_l - R_{\text{ref}})$ represents the maximum distance of the sensing targets of interest. In addition, the maximum BF of the desired IF signal must satisfy 
 \begin{equation}
 \small
 \label{IFMAX}
    \frac{2 \vert\Delta R_{\text{max}}\vert}{\nu} <  \frac{N_{\text{G}}^{\text{S}}T}{N} .
 \end{equation}  
Thus, there is no mis-mixing between the SPS and PS. From (\ref{IFMAX}), the maximum detectable range can be extended by:  1) increasing the guard-band size $N_{\text{G}}^{\text{S}}$,  2) reducing $\Delta f$, and 3) selecting a reference distance $R_{\text{ref}}$ closer to the region of interest. In fact,  $R_{\text{ref}}$ serves the midpoint of the sensing window, and the proposed system can detect targets within $\left (R_{\text{ref}}-\frac{\nu N_{\text{G}}^{\text{S}}T}{2N}, R_{\text{ref}}+\frac{\nu N_{\text{G}}^{\text{S}}T}{2N} \right)$.

\textit{Remark 3:  For monostatic sensing, in the presence of on-board leakage from the ISAC transmitter, mixing the SPS with its local reference produces a direct current component, whereas the other subcarriers generate IF components at much higher BFs that can be effectively suppressed by the LPF. By placing the ADC after the LPF, saturation induced by transmitter leakage can be mitigated, thereby eliminating the need for a complex and costly full-duplex design. Also, since the ISAC transmitter and radar receiver are co-located and share the same LO, carrier frequency synchronization is inherently guaranteed and the correlated phase noise is largely canceled during the dechirping process.}

\textit{Remark 4: As the maximum BF of the desired IF signal is significantly lower than the bandwidth, i.e., $\text{BF}_{\frac{N}{2}, \text{max}}\ll B$, the ADC sampling rate can be substantially reduced relative to the original signal bandwidth, thereby lowering the hardware complexity.   Compared to conventional communication-centric sensing receivers that require a wideband ADC with sampling rate $\geq B$, full-duplex SIC, and digital-domain matched filtering over the entire signal bandwidth, the proposed receiver relies only on a low-cost analog mixer, an LPF, and a narrowband ADC, achieving a substantial reduction in both hardware cost and computational complexity. }

\color{black}

\subsection{ AFDM Parameter-Oriented Estimation}
\label{Secest}
After the LPF, the resultant desired signal is further digitalized by a low sampling-rate ADC. By denoting  $f_{\text{ADC}}$ as the ADC sampling rate,  the output signal is given by 
\begin{equation}
\begin{aligned}
\label{adc_sig}
\small
     &   r_{\text{ADC},i}^k[m] \\
       & \simeq   \sum_{l=0}^{L-1} \left [  \sqrt{{P}_{\text{SPS}}}h_l^{\text{S}}  e^{ \text{IF}_{\frac{N}{2},i}^{k}  (m\Delta t_s^k)}  \!+ \! w(m\Delta t_s^k) \right ]  \Large \vert _{\Delta t_s^k = \frac{1}{f_{\text{ADC}}}}  \\
       & \simeq  \! \sum_{l=0}^{L-1} \! \sqrt{{P}_{\text{SPS}}} h_l^{\text{S}}  e^{\!   -j2\pi  \frac{\frac{2\alpha \Delta R }{\nu} \!+\! f_d^0}{f_{\! \text{ADC}}}m } e^{\! \! -j2\pi  f_d^0 T_{\text{s}}i} e^{\!- \!j2\pi  \frac{{\phi}^0_i}{f_{\text{ADC}}}} \!\!+ \!\!w_i^k[m].
\end{aligned}
\end{equation}

\begin{figure}
    \centering
\includegraphics[width=0.88\linewidth]{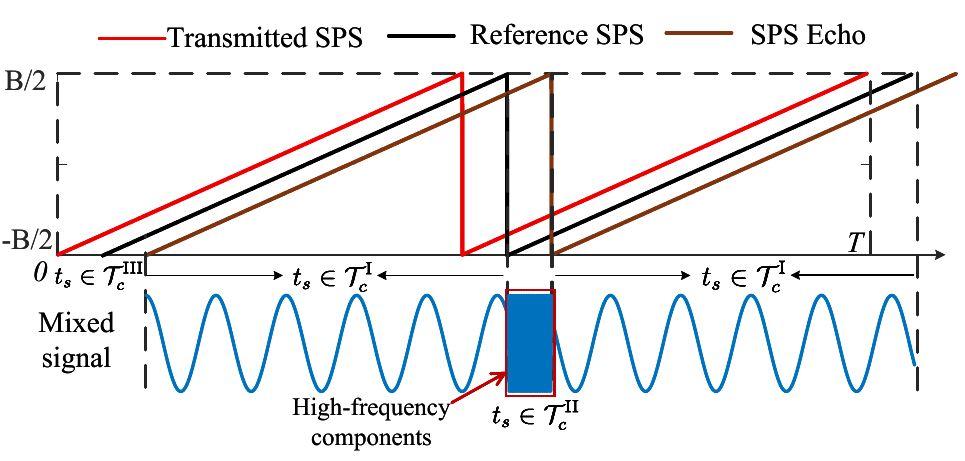}
    \caption{ Illustration of the  transmitted SPS, the corresponding received echo, and the IF signal  with $K=2$.}
    \label{SPSecho}
    \vspace{-0.6em}
 \end{figure}
 
Before presenting the AFDM parameter-oriented radar processing schemes, we discuss the phase-hopping issue in the IF signal (i.e., (\ref{IFafdm}) and  (\ref{adc_sig})) caused by spectrum wrapping. 
Define the IF signals of the $k$th and $(k+1)$th components in (\ref{IFafdm}) to be effectively continuous if   $\text{IF}_{\frac{N}{2},i}^{k} (t) =  \text{IF}_{\frac{N}{2},i}^{k+1} (t)$, $ t =kT/K $.  Based on (\ref{IFafdm}), the above equality   holds when 
\begin{equation}
\small
\label{phaseeq}
    {  \frac{\Delta R_lB}{\nu}  +\frac{ v_lB(1+kT/K)}{\nu}  }  \in \mathbb Z.
\end{equation} 
(\ref{phaseeq}) is a stringent requirement since $\Delta R_l$ and $v_l$ are unknown parameters that need to be estimated. Fig.~\ref{SPSecho} illustrates an example of the IF signal with $c_1 = \frac{1}{N}$ where (\ref{phaseeq}) does not hold. Although the undesired IF signal within the hopping window, i.e., $t_s \in \mathcal T_k^{\text{II}}$, can be suppressed by the LPF, phase discontinuities still arise between adjacent chirp components within each AFDM symbol. Such discontinuities (or hopping) pose significant challenges for radar parameter estimation. For example, they may cause increased sidelobes and spectral leakage when on-grid estimation methods, such as FFT-based techniques, are applied for range estimation \cite{ranasinghe2024joint}. To address this issue, we propose AFDM-parameter-oriented sensing algorithms.

\begin{figure}
    \centering
    \includegraphics[width=1\linewidth]{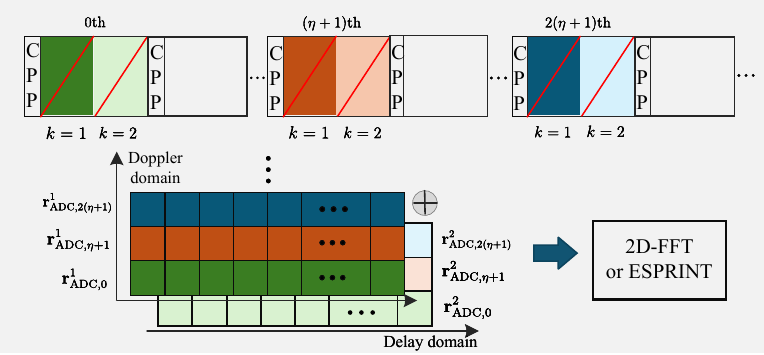}
    \caption{Illustration of the proposed AFDM-oriented parameter estimations.}
    \label{FusiinAlg}
    \vspace{-.5em}
\end{figure}

From (\ref{IFafdm2}), it can be observed that the Doppler term 
remains constant across the ISAC symbol index $i$ 
and the chirp sweeping index $k$. This indicates  Doppler and BF can be estimated based on the $k$th IF component  across different ISAC symbols.  Moreover, it is noted that the terms 
${f}_d^0$ and ${\text{BF}}_{\frac{N}{2},i}^0$ in 
\eqref{IFafdm2} are also independent of the chirp sweeping 
index $k$. This motivates us to aggregate all $K$ chirp 
components within each ISAC symbol, and 
then employ the aggregated signals across different ISAC 
symbols for joint range and velocity estimation, as 
illustrated in Fig.~\ref{FusiinAlg}. Specifically, the $K$ chirp components within the $i$th 
symbol are aggregated as follows:
\begin{equation}
\small
    \mathbf{r}_{\text{agg},i}[m] = \sum_{k=0}^{K-1} 
     r_{\text{ADC},i}^{k}[m].
\end{equation}
Then, by collecting the $N_s$ ISAC symbols into 
$\mathbf{R}_{\text{agg}} = 
[\mathbf{r}_{\text{agg},0}^{\mathcal{T}}, 
\mathbf{r}_{\text{agg},1}^{\mathcal{T}}, \ldots, 
\mathbf{r}_{\text{agg},N_s-1}^{\mathcal{T}}]$, 
existing algorithms such as 2D-FFT and ESPRIT can be applied 
to estimate $f_d^0$ and the beat frequency 
$\frac{\frac{2\alpha \Delta R_l}{\nu} + f_d^0}{f_{\text{ADC}}}$ 
\cite{ranasinghe2024joint,ESPRIT1989}.  


%

\subsection{Performance Indicators}
We now present the performance indicators of the proposed AFDM-ISAC scheme.

\subsubsection{Maximum velocity and velocity resolution}
From (\ref{IFafdm}), the initial phase difference of the $k$th chirp component between two consecutive AFDM-ISAC symbols is
 $    \Delta \Phi \triangleq \text{IF}_{\frac{N}{2},i+1}^{k} (0) -  \text{IF}_{\frac{N}{2},i}^{k} (0)    \approx \frac{4\pi v_lT_{\text{s}}f_c  }{\nu}$.
 The speed measurement is unambiguous only if $\Delta \Phi < \pi$. Hence,  the maximum speed is given by 
\begin{equation}
 \small
 \label{vmax}
    v_{\text{\text{max}}} =    \frac{\nu}{4 f_c(1+\eta)(T + T_{\text{CPP}})}.
\end{equation}


  According to the Rayleigh limit, two Doppler components are resolvable when their frequency separation satisfies $\Delta f_d \geq \frac{1}{N_{\text{F}}   (T + T_{\text{CPP}})}$, leading    to  the Rayleigh-limited velocity resolution as follows: 
\begin{equation}
\small
\label{vres_ray}
    v_{\text{res}}^{\text{Ray}} =    \frac{\nu}{2N_{\text{F}} f_c(T + T_{\text{CPP}})}.
\end{equation}
It is worth noting that (\ref{vres_ray}) represents a fundamental limit for classical spectral estimation methods such as the FFT-based algorithms. In practice, super-resolution algorithms, such as ESPRIT \cite{ESPRIT1989} adopted in this paper, can resolve targets with velocity separations well below the Rayleigh limit \cite{10423051}. Specifically, the achievable velocity resolution of ESPRIT-type algorithms is governed by the estimation accuracy, i.e.,  the CRLB. In this paper, the CRLB-limited velocity resolution  is essentially the estimation of multiple tone frequencies, which can be approximated as \cite{stoica1989music}
\begin{equation}
\small
\label{vres_sr}
    v_{\text{res}}^{\text{SR}} \approx \frac{\nu}{2\pi f_c ( 1+\eta)(T + T_{\text{CPP}})} \sqrt{\frac{6}{N_{\text{s}}(N_{\text{s}}^2-1) \cdot \text{SNR}}},
\end{equation}
where $\text{SNR}$ refers to  the received signal-to-noise ratio (SNR).

\color{black}

\subsubsection{Maximum range and range resolution} The system's maximum range depends on the guard-band length (i.e., $N_{\text{G}}^{\text{S}}$), the reference delay (i.e., $\tau_{\text{ref}}$), and the ADC sampling rate. Following the Nyquist theorem, the ADC rate must be at least twice the maximum BF. From (\ref{IFMAX}), it follows that
\begin{equation}
\small
\label{Rmax}
    R_{\max} = \min \Big \{      \frac{N_{\text{G}}^{\text{S}}T}{2N} + R_{\text{ref}}, \quad  \frac{f_{\text{ADC}}\nu}{4\alpha} +  R_{\text{ref}} \Big \}.
\end{equation}

A straightforward approach for  extending the maximum detection range is to increase the number of sensing guard subcarriers $N_{\text{G}}^{\text{S}}$, which, however, comes at the cost of reduced communication SE. In contrast, (\ref{Rmax}) reveals that by introducing a reference delay to the local chirp signal, as defined in (\ref{reflochirp}), the maximum detection range can be flexibly enlarged by simply increasing $\tau_{\text{ref}}$, without sacrificing any communication resources.

To resolve two closely spaced targets in range, the spacing between their BF tones must exceed the Rayleigh limit, i.e., $\Delta f_{\text{IF}} \geq \frac{1}{T_{\text{obs}}}$, where $T_{\text{obs}}$ denotes the observation time of the IF signal. Since the observation time satisfies $T_{\text{obs}} \leq \frac{T}{K}$ and the BF is related to range by $f_{\text{IF}} = \frac{2\alpha R}{\nu}$, the Rayleigh-limited range resolution is given by
\begin{equation}
\small
\label{Rres_ray}
    R_{\text{res}}^{\text{Ray}} = \frac{\nu}{2\alpha T_{\text{obs}}}  \leq \frac{K\nu}{2\alpha T}.
\end{equation}

Similar to the velocity case,  the 
 range resolution of super-resolution-based algorithms can be approximated as  \cite{stoica1989music}
\begin{equation}
\small
\label{Rres_sr}
    R_{\text{res}}^{\text{SR}} \approx \frac{\nu}{2\pi \alpha / f_{\text{ADC}}} \sqrt{\frac{6}{ M_{\text{ADC}}(M_{\text{ADC}}^2-1) \cdot \text{SNR} }},
\end{equation}
 where  $M_{\text{ADC}} = \lfloor f_{\text{ADC}} T_{\text{obs}} \rfloor$ is the number of ADC samples.

\subsubsection{Normalized communication efficiency}  The normalized communication efficiency (NCE) is defined as  
\begin{equation}
    \small
    \label{NCE}
    \begin{aligned}
            \eta_{\text{NCE} }&=  \frac{\text{Number of communication subcarriers}}{\text{Number of total subcarriers}} \\
   &= \frac{N-2N_{\text{G}}^{\text{D}}-2N_{\text{G}}^{\text{S}}-3+ \eta N}{(1+\eta)N} \\
   & = 1- \frac{1}{1+\eta}\cdot \frac{2N_{\text{G}}^{\text{D}}+2N_{\text{G}}^{\text{S}}+3}{N}.
    \end{aligned}
\end{equation}
From (\ref{NCE}), it can be observed that the NCE is dominated by the parameter $\eta$ and the number of guard subcarriers.   A larger $\eta$ means higher communication SE.     However, increasing $\eta$ comes at the cost of reducing the maximum unambiguous detectable speed $v_{\max}$, as given in (\ref{vmax}). To illustrate this trade-off, we consider a typical parameter setting with $\Delta f = 15$ kHz, $N = 256$, and a CPP overhead ratio of $0.09$. The corresponding maximum detectable speeds for $\eta = \{0, 1, 2, 3, 4\}$ are $v_{\max} = \{925, 462, 308, 231, 185\}$ km/h, respectively. It can be observed that even when $\eta = 4$, the system can still support a maximum speed of $185$ km/h. This indicates that $\eta$ has a   large degree of freedom for flexible selection, allowing the system to achieve both high communication SE and satisfactory sensing performance requirements simultaneously.
 \color{black}

\section{The Proposed Channel Estimation Schemes}

Based on the proposed AFDM-ISAC frame structure, this section studies efficient channel estimation methods. We employ the  BEM  to characterize time-varying channels due to its simplicity and analytical tractability \cite{liu2022near}.   Specifically, we begin with a GCE-BEM–assisted estimator, followed by a power-allocation strategy and a KF-BEM channel estimation method tailored to the frame-based design.

\subsection{ The GCE-BEM }

In this paper,  we employ  the GCE-BEM   to model $h_i(n,p)$ in (\ref{DFIR}) \cite{liu2022near}, i.e.,
   \begin{equation}
   \small
   \label{h_ti}
       {h_{i}}\left ({{n,p} }\right) = \sum  _{q = 0}^{Q - 1}  {b_{n }^q}c_{i,p}^{q},
   \end{equation} where ${b_{n}^q} = {{\mathrm{e}}^{\frac {j2\pi (q - Q/2)n}{NR}}}$ denotes the $q$th basis coefficient at the $n$th entry, $c_{i,p}^{q}$ denotes the corresponding  BEM coefficient at the $i$th symbol of the $p$th tap, and $Q$ is the dimension of the basis vector that satisfies $Q \ge 2\left\lceil {R f_{\text{d,max}} N T_s} \right\rceil$  with $f_{\text{d,max}}$ denoting  the maximum Doppler frequency (Hz) and $R$ representing a  positive integer resolution parameter.   In this paper  $R = 2$ is considered as it yields sufficiently low error, and  $R \geq 3$  provides negligible benefit, as shown in   \cite{liu2022near}.

Let $\mathbf b^q = [b_0^q, b_1^q, \ldots, b_{N-1}^q]^{\mathcal T}$,   $\textbf {c}^{q}_i=[c_{i,0}^q, c_{i,1}^q, \ldots, c_{i,P-1}^q]^{\mathcal T}$   and $\mathbf {B}^{q} = \text {diag}\{\mathbf {b}^{q}\} $.   Then, the time-domain channel can be expressed in  matrix form as 
\begin{equation}
\small
\label{Hi_2}
       {\mathbf {H}}_i \!  =  \! \sum \limits _{q = 0}^{Q-1} {\sum \limits _{p = 0}^{P-1} {c_{i,p}^q}{{{\mathbf {B}}^{q}}{{\boldsymbol {\Pi }}^{p}}} }   \! \overset{(\text{i})}{=}\! \sum _{q=0}^{Q-1} \mathbf {B}^{q} \textbf {F}_{N}^{\mathcal H}\text {diag}\{\textbf {F}_{N\times P}\textbf {c}^{q}_i\}\textbf {F}_{N}, 
\end{equation}
where  $\boldsymbol {\Pi }$ denotes a  circulant matrix,   and (i) is obtained with the aid of DFT, i.e., 
 $  {\sum \nolimits _{p = 0}^{P-1} {c_{i,p}^q}{ {{\boldsymbol {\Pi }}^{p}}} }    =\textbf {F}_{N}^{\mathcal H}\text {diag}\{\textbf {F}_{N\times P}\textbf {c}^{q}_i\}\textbf {F}_{N},$  where $\textbf {F}_{N\times P}$ denotes the first $P$ columns of the  DFT matrix  $\textbf {F}_{N}$.    
 
Substituting (\ref{Hi_2}) into (\ref{yi_1}),    the  received signal with BEM  can be expressed as 
\begin{equation}
\small
\label{Hi}
\begin{aligned}
        \mathbf y_i & =    \sum_{q=0}^{Q-1}   \mathbf A    \mathbf B^q \textbf {F}^{\mathcal H}\text {diag}\{\textbf {F}_{N\times P}\textbf {c}^{q}_i\} \textbf {F}   \mathbf A^{\mathcal H}  \mathbf {x}_i  + \mathbf n_i \\
         & =   \sum_{q=0}^{Q-1}   \mathbf A    \mathbf B^q \textbf {F}^{\mathcal H}\text {diag}\{\textbf {F} \mathbf A^{\mathcal H}  \mathbf   {x}_i \} \mathbf {F}_{N\times P} \textbf {c}^{q}_i+\mathbf n_i.
\end{aligned}
\end{equation}
 We  denote   $ \mathbf {x}_{i, \text{p}}$ and $ \mathbf {x}_{i, \text{d}}$,  $ i \in \mathcal I_{\text{ISAC}}$  as the associated pilot and data sections of the $i$th  symbol, respectively.  Then,  (\ref{Hi})   can be rewritten as    
\begin{equation}
\small
\label{yii}
\begin{aligned}
        \mathbf y_i   & =        \sum_{q=0}^{Q-1}   \mathbf A    \mathbf B^q \textbf {F}^{\mathcal H}\text {diag}\{\textbf {F} \mathbf A_{\text{p}}^{\mathcal H}  \mathbf {x}_{i, \text{p}}\} \mathbf {F}_{P} \textbf {c}^{q}_i+ \\ 
         & \quad  \sum_{q=0}^{Q-1}   \mathbf A    \mathbf B^q \textbf {F}^{\mathcal H}\text {diag}\{\textbf {F} \mathbf A^{\mathcal H}_{\text{d}}  \mathbf {x}_{i, \text{d}}\} \mathbf {F}_{P} \textbf {c}^{q}_i  + \mathbf n_i  \\
       & \equiv  \mathbf M_{i, \text{p}} \textbf {c}_i+   \mathbf M_{i, \text{d}} \textbf {c}_i  + \mathbf n_i,   i \in \mathcal I_{\text{ISAC}},
\end{aligned}
\end{equation}
where $\mathbf M_{i, \text{p}}  = [   \mathbf M_{i, \text{p}}^0, \mathbf M_{i, \text{p}}^1, \ldots,  \mathbf M_{i, \text{p}}^{Q-1}  ]$, $ \mathbf M_{i, \text{p}}^q =  \mathbf A    \mathbf B^q \textbf {F}^{\mathcal H}\text {diag}\{\textbf {F} \mathbf A^{\mathcal H}_{\text{p}}  \mathbf {x}_{i, \text{p}}\} \mathbf {F}_{P}$, $\mathbf M_{i, \text{d}}  = [   \mathbf M_{i, \text{d}}^0, \mathbf M_{i, \text{d}}^1, \ldots,  \mathbf M_{i, \text{d}}^{Q-1}  ]$, $ \mathbf M_{i, \text{d}}^q =  \mathbf A    \mathbf B^q \textbf {F}^{\mathcal H}\text {diag}\{\textbf {F} \mathbf A^{\mathcal H}_{\text{d}}  \mathbf {x}_{i, \text{d}}\} \mathbf {F}_{P}$
and $\textbf {c}_i = [\textbf {c}_i^{0,{\mathcal T}}, \textbf {c}_i^{2,{\mathcal T}} \ldots, \textbf {c}_i^{Q-1} ]^{\mathcal T}$. In the following, the matrices $\mathbf M_{i, \text{p}} $ and $\mathbf M_{i, \text{d}} $ are referred to as measurement matrices.   In addition, denote $\mathbf T_\text{p} = [\mathbf I_{N}]_{\mathcal{N}_{\text{p}}}$ as  the sub-matrix of $\mathbf I_{N}$ obtained by selecting  the  rows indexed by  ${\mathcal{N}_{\text{p}}}$, where $ \mathcal{N}_{\text{p}} =  \{ n \vert n \in [\frac{N}{2}-N_{\text{G}}^{\text{D}}-N_{\text{G}}^{\text{S}}-1,  \frac{N}{2}+N_{\text{G}}^{\text{D}}+1] \} $ is the  index set of the pilot sections. Multiplying $\mathbf T_\text{p}$ on both  sides of (\ref{yii}) leads to 
\begin{equation}
\small
    \mathbf y_{i, \text{p}} =  \widetilde{  \mathbf M}_{i, \text{p}}  \textbf {c}_i+   \underbrace {\widetilde{ \mathbf M}_{i, \text{d}} \textbf {c}_i  + \widetilde{\mathbf n}_i}_{\text{Interference}},
\end{equation}
where $\mathbf y_{i, \text{p}} = \mathbf T_\text{p}\mathbf y_{i}$   denotes the received pilot   section,   $  \widetilde{  \mathbf M}_{i, \text{p}}  =  \mathbf T_\text{p}\mathbf M_{i, \text{p}} $, $  \widetilde{  \mathbf M}_{i, \text{d}}  =  \mathbf T_\text{p}\mathbf M_{i, \text{d}} $,  $  \widetilde{ \mathbf {n}}_{i }   =  \mathbf T_\text{p}\mathbf {n}_{i }  $ denote  the transmitted pilot section, the interference from the data subcarriers and the corresponding noises term, respectively. Finally, the  linear minimum mean square error (LMMSE) estimator is applied to obtain the BEM coefficients, i.e., 
\begin{equation}
\small
\label{mmse}
    \mathbf W_i^\text{LMMSE} = \mathbf R_{\text{c}}  \widetilde{  \mathbf M}_{i, \text{p}} ^{\mathcal H} \left( \widetilde{  \mathbf M}_{i, \text{p}}  \mathbf R_{\text{c}}   \widetilde{  \mathbf M}_{i, \text{p}} ^{\mathcal H} + \mathbf R_{\text{d}} +  \mathbf R_{\text{n}} \right)^{-1},
\end{equation}
where $\mathbf R_{\text{c}} = \mathbb E\{ \mathbf c \mathbf  c^{\mathcal H} \}$,  $  \mathbf R_\text{d}   = \mathbb E \{\widetilde{ \mathbf M}_{i, \text{d}} \textbf {c}_i    \mathbf {c}_i^{\mathcal H} (\widetilde{ \mathbf M}_{i, \text{d}}  )^{\mathcal H}\} \}$  and $\mathbf R_{\text{n}} = \sigma_c \mathbf I_{\vert\text{Ind}_{\text{p}}\vert}$ are the covariance  matrices.  The detailed derivations of    $\mathbf R_{\text{c}}$  and $  \mathbf R_{\text{d}}$ are given in  Appendix \ref{eqRcd}.
 With (\ref{mmse}), the  BEM coefficient vector  is  estimated by
\begin{equation}
\small
\label{bem_che}
     \widehat{ \mathbf c}_i =  \mathbf W_{\text{LMMSE}}^{i} \mathbf y_{i,\text{p}},   i \in \mathcal I_{\text{ISAC}}. 
\end{equation}
Finally, the estimated   affine-domain channel   is   given by 
\begin{equation}
\small
\label{channle_H}
\begin{aligned}
    \widehat{ \mathbf H}_{\text{eff},i}   =  \mathbf A   \sum_{q=0}^{Q-1} \mathbf B^q \textbf {F}^{\mathcal H}\text {diag}\{\textbf {F}_{N\times P} \widehat{ \mathbf c}^{q}_i\} \mathbf F,   i \in \mathcal I_{\text{ISAC}}.
\end{aligned}
\end{equation}

\subsection{Power Allocation }

Given the proposed frame structure and channel estimation scheme, it is important  to investigate the    power allocation between pilot and data symbols under a constrained total power budget, as defined in (\ref{xn}). 
Denote the channel estimation error matrix in the DAFT domain by $  \widehat{ \mathbf H}_{\text{err},i}   = { \mathbf H}_{\text{eff},i} -  \widehat{ \mathbf H}_{\text{eff},i}$. Using MMSE detector, the estimated symbol vector is given as 
\begin{equation}
\small
\label{equa}
\begin{aligned}
         \widehat{ \mathbf x}_i       \! =\!{\mathbf G_{\text{LMMSE}}^{i}}  \mathbf y_{i} \!= \!  \mathbf R_{{ \mathbf x}_i } \widehat{ \mathbf H}_{\text{eff},i}^{\mathcal H}\Big(   \widehat{ \mathbf H}_{\text{eff},i}  \mathbf R_{{ \mathbf x}_i } \widehat{ \mathbf H}_{\text{eff},i}^{\mathcal H}  + \mathbf R_{\text{I}} \Big)^{-1}  \mathbf y_{i}.
\end{aligned}
\end{equation}
To derive the  SINR on the $n$th chirp subcarrier, 
we rewrite the received signal as 
\begin{equation}
\small
    \mathbf y_{i} =  \widehat{ \mathbf H}_{\text{eff},i}\mathbf x_i   + \underbrace{\widehat{ \mathbf H}_{\text{err},i}\mathbf x_i    + { \mathbf n}_i}_{\text{Interference+ noise}}.
    \label{recy}
\end{equation}
  Then  (\ref{equa}) can be further written as 
\begin{equation}
\small
\begin{aligned}
      \widehat{   \mathbf x}[n] = & \mathbf T_{\text{est},i}[n,n] \mathbf x[n] \\
      & + \sum_{m \neq n}  \mathbf T_{\text{est},i}[n,m] \mathbf x[m] +   \mathbf T_{\text{err},i} [n,n]+   \mathbf T_{\text{n},i} [n,n],
\end{aligned}
\end{equation}
 where $\mathbf T_{\text{est},i} = {\mathbf G_{\text{LMMSE}}^{i}} \widehat{ \mathbf H}_{\text{est},i}  $,   $\mathbf T_{\text{err},i} = {\mathbf G_{\text{LMMSE}}^{i}} \widehat{ \mathbf H}_{\text{err},i}  $,    $\mathbf T_{\text{n},i} = {\mathbf G_{\text{LMMSE}}^{i}} \widetilde{\mathbf n}_i  $.  Assuming the transmitted data   has the unit power, the SINR on the $n$th subcarrier   can be expressed as 
 \begin{equation}
 \small
     \text{SINR}_n\!\! = \!\! \frac{ \mathbf T_{\text{est},i}^2 [n,n] }{\text{Var}\Big \{ \!     \sum\limits_{m \neq n} \!\! \mathbf T_{\text{est},i}[n,m] \mathbf x[m] \! +\!   \mathbf T_{\text{err},\! i} [n,\!n] \!+\!  \mathbf T_{\text{n},i} [n,\!n]   \!  \Big  \}}. 
     \label{SINR}
 \end{equation}
By using the central limit theorem and based on the results reported in \cite[ (15)-(16)]{li2024chirp}, the equality in (\ref{SINR}) can be approximated as $  \text{SINR}_n  \approx \frac{ \mathbf T_{\text{est},i}  [n,n] }{1-\mathbf T_{\text{est},i}  [n,n]}$.
Note that the power of the SPS is determined by radar sensing requirements; thus, only $P_{\text{SP}}$ needs to be optimized. Hence, the power allocation problem is formulated as
\begin{equation}
    \small 
    \label{SINRn}
    \begin{aligned}
            \underset{P_{\text{PS}}}{\max}  \quad \quad & \text{SINR}_n,  n \in \mathcal N_{\text{D}} \\
 \text{s.t.}   \quad \quad &  P_{\text{SPS}} + 2P_{\text{SP}} + \vert \mathcal N_{\text{D}} \vert =  {P_{\text{total}}},
    \end{aligned}
\end{equation}
where each data symbol is assumed to have unit average power, and $P_{\text{total}}$   denotes   the total available power. Once $\text{SINR}_n$ for each data subcarrier   is obtained, the average bit error rate (BER) performance can be derived by following the approach in \cite[ (16)--(21)]{li2024chirp}. With the statistical channel information, (\ref{SINRn}) can be resolved via Monte Carlo simulation.


\subsection {KF-BEM  Enhanced  Channel Estimation }
 Note that once the estimated channel is obtained for the ISAC symbol based on (\ref{bem_che}) and (\ref{channle_H}), the effective channel for the subsequent pure data symbols can be predicted using (\ref{hsv}), i.e.,  $\mathbf h_{i+1} =  \widetilde{ \mathbf S} \mathbf h_{i}$. While the state transition matrix $\widetilde{\mathbf S}$ enables channel prediction, $\mathbf h_{i}$ is only available through noisy observations with random dynamics, making the prediction inaccurate and prone to error propagation. Hence, we propose a KF-BEM scheme to optimally fuse both predictions and observations, thereby achieving robust channel predication under the proposed frame-based structure.
 The proposed KF-BEM scheme   mainly consists of 1) State space model of BEM coefficients; 2) State prediction; and 3) Sate updating.

\subsubsection{State space model of  BEM coefficients}

Using $ \mathbf h_{i} =  (\mathbf I_P \odot \mathbf B) \mathbf c_{i}$ and   (\ref{hsv}),  the state space of  BEM coefficients can be expressed as 
\begin{equation}
\small
\label{StatePSpace}
\begin{aligned}
        \mathbf c_{i+1} & =  \mathbf S \mathbf  c_{i} + \mathbf v_i,  \\
        \mathbf y_i &  = \mathbf M_i \mathbf c_{i} + \mathbf n_i,
\end{aligned}
\end{equation}
 where $  \mathbf S =  \left (( \mathbf I_P \odot \mathbf B)^{\mathcal H}  (\mathbf I_P \odot \mathbf B) \right)^{-1}   (\mathbf I_P \odot \mathbf B) \widetilde{ \mathbf S}(\mathbf I_P \odot \mathbf B) $ and  $ \mathbf v_i = \left (( \mathbf I_P \odot \mathbf B)^{\mathcal H}  (\mathbf I_P \odot \mathbf B) \right)^{-1}   (\mathbf I_P \odot \mathbf B)  \mathbf n _i$.  $\mathbf M_{i}$ is  the measurement matrix of the transmitted data, i.e,    $\mathbf M_{i}  = [   \mathbf M_{i}^0, \mathbf M_{i}^1, \ldots,  \mathbf M_{i}^Q  ]$  and $ \mathbf M_{i}^q =  \mathbf S    \mathbf B^q \textbf {F}^{\mathcal H}\text {diag}\{\textbf {F} \mathbf A^{\mathcal H}  \mathbf {x}_{i}\} \mathbf {F}_{P}$.

\subsubsection{State prediction} In the state prediction process, \textit{a priori} estimates of the state variable at the $(i+1)$th moment are obtained based on the  \textit{a posterior} estimates and the state transfer function. The state prediction of the state space model in  (\ref{StatePSpace}) can be obtained as 
    \begin{align}
    \small
    \label{eqc}
        {\mathbf c}_{i|i - 1} & = {\mathbf S} {\mathbf c}_{i - 1},  \\
                {\mathbf P}_{i|i - 1}& = {\mathbf S} {\mathbf P}_{i - 1} {\mathbf S}^{\mathcal H} + \mathbf V,  
    \end{align}
 where ${\mathbf P}_{i|i - 1}$ and ${\mathbf P}_{i - 1} $ denote the   \textit{a priori}  and  \textit{a posterior}  covariance matrices of the $i$th state  variable, and $\mathbf{V}   = \mathbb{E}[\mathbf{v}_i \mathbf{v}_i^{\mathcal{H}}]  =   \sigma_n^2     \mathbf{I}_{PQ} $ denotes the covariance matrix of  $\mathbf v_i$.
 Note that for the data symbols, the measurement matrix $\mathbf M_{i}$  is unknown and contains the transmitted data symbols. Hence, we propose a  refined channel estimation  scheme to construct the  $\mathbf M_{i}$.  Specifically, based on (\ref{eqc}),  \textit{a priori}  estimate of BEM coefficients $  {\mathbf c}_{i|i - 1}$ is first obtained, and then the estimated channel, denoted as $ \widehat{ \mathbf H}_{\text{eff},i|i-1}  $, can be obtained based on (\ref{channle_H}). Therefore, the  $i$th transmitted AFDM symbol can be obtained by LMMSE equalization as
 
\begin{equation}
\small
\label{e_est_}
   \widehat{\mathbf x}_i =  \left(   \widehat{ \mathbf H}_{\text{eff},i|i-1} \widehat{ \mathbf H}_{\text{eff},i|i-1}  + \sigma_{{c}} \mathbf I _N\right)^{-1}  \widehat{ \mathbf H}_{\text{eff},i|i-1}^{\mathcal H} \mathbf y_{i},
\end{equation}
\color{black}
where $\widehat{\mathbf x}_i$ denotes the predicted transmitted symbols. As the transmitted   symbols  are chosen from the constellation alphat set $\boldsymbol{\mathcal X} = \{\mathcal X_1,  \mathcal X_2, \ldots, \mathcal X_M\}$, the output of the refined channel estimation  is a modulation symbol in $\boldsymbol{\mathcal X}$ which is the nearest one for $ \widehat{\mathbf x}_i [n]$., i.e., $ \widehat{\mathbf x}_i  [n] = \underset{\mathcal X_m \in \boldsymbol{\mathcal X}}{\min} \Vert \mathcal X_m  - \widehat{\mathbf x}_i [n]\Vert^2$. Finally, the measurement matrix $\widehat {\mathbf M}_{i}$  can be constructed from $\widehat{\mathbf x}_i   $  and  will be used for  state updating. 

\subsubsection{State updating}  After the state predication, the  \textit{a posterior} state BEM coefficients are estimated through  the state updating principles of Kalman filter as follows:
\begin{equation}
\small
    \label{eqsu}
     \begin{aligned}
     {\mathbf K}_{i} & =   {\mathbf P}_{i|i - 1}   \widehat {\mathbf M}_{i}^{\mathcal H} \big( \widehat {\mathbf M}_{i}{\mathbf P}_{i|i - 1}  \widehat {\mathbf M}_{i}^{\mathcal H} + \mathbf V  \big)^{-1}, \\
                {\mathbf c}_{i}& = {\mathbf c}_{i|i - 1} +  {\mathbf K}_{i}    \big(  \mathbf y_i-\widehat {\mathbf M}_{i} {\mathbf c}_{i|i-1}\big),\\
                {\mathbf P}_{i}& = {\mathbf P}_{i|i - 1}-    {\mathbf K}_{i}   \widehat {\mathbf M}_{i} {\mathbf P}_{i|i - 1},
    \end{aligned}
\end{equation}
where ${\mathbf K}_{i} $ denotes the Kalman gain. Based on the state prediction and updating of the Kalman filter, the \textit{a posterior}   BEM coefficient $  {\mathbf c}_{i}$  is obtained as the output, and the effective channel $ \widehat{ \mathbf H}_{\text{eff},i}$ is obtained  based on (\ref{channle_H}).  Finally, the  detailed steps of the proposed KF-BEM are summarized in \textbf{Algorithm \ref{EF_BEM}}.

 \begin{algorithm}[t] 
\caption{The Proposed KF-BEM for Frame-based Channel Estimation.}
\begin{algorithmic}[1]
    \STATE Initialize KF-BEM.
        \STATE \textbf{Step 1: }  If $\mod(i,\eta+1)=0$,  the BEM coefficients and channel matrix are estimated by the ISAC symbol based on (\ref{bem_che}) and (\ref{channle_H}), respectively.  When the next ISAC symbol arrived, \textbf{Step 2} to  \textbf{Step 4}  are executed. 
        \STATE \textbf{Step 2: }  State Prediction and Measurement Matrix Construction:   The \textit{a priori} estimates of the state variable $    {\mathbf c}_{i|i - 1}$ is calculated through (\ref{eqc}). Then, the measurement matrix $\widehat {\mathbf M}_{i}$  can be constructed based on the resultant $\widehat{\mathbf x}_i $ in   (\ref{e_est_}). 
        \STATE 
\textbf{Step 3: } State Updating: The  \textit{a posterior} state BEM coefficients are updated by (\ref{eqsu}). Then, the effective channel of the $i$th symbol is obtained by (\ref{channle_H}) based on the $\mathbf c_i$ in (\ref{eqsu}).
    \STATE   \textbf{Step 4: }   Set $i = i + 1$. If  $\mod(i,\eta+1)\neq 0$,  repeat \textbf{Step 2 } and \textbf{Step 3}, otherwise jump back to \textbf{Step 1}. 
\end{algorithmic}
\label{EF_BEM}
\end{algorithm}

\subsection{Benchmarking Schemes and Complexity Analysis}  
In this paper, the existing embedded channel estimation (ECE) scheme from \cite{MIMOAFDMchannest}, referred to as diagonally reconstructed ECE (DR-ECE), is employed as a benchmark. Since the original DR-ECE in \cite{MIMOAFDMchannest} utilizes only a single pilot for channel estimation, we modify it here to accommodate the proposed AFDM-ISAC structure.   
In the modified DR-ECE scheme, the threshold-based magnitude detection on the multiple received pilots is first conducted, i.e.,  
\begin{equation}
\label{drece}
\begin{aligned}
         {\widehat{\mathbf h}}_{\text{eff},i}^{(m)}[n] = \begin{cases}   {\mathbf y}_{\text{p},i}^{(m)}[n]/x_{\text{pilot}}^{(m)},  &  \sum_{m=1}^{3} \vert{\mathbf y}_{\text{p},i}^{(m)}[n]\vert > \gamma_{\text{th}},    \\
     0, & \text{otherwise},
     \end{cases}
\end{aligned}
\end{equation}
where $   {\widehat{\mathbf h}}_{\text{eff},i}^{(m)}  \in \mathbb C^{(QP+1) \times 1}$ and ${\mathbf y}_{\text{p},i}^{(m)}  \in \mathbb C^{(QP+1) \times 1} $ denote the effective channel and received data associated with the $p$th pilot, respectively \cite{MIMOAFDMchannest}.
The  DR-ECE scheme rely on the integer delay to reconstruct  the channel. For multiple pilots, the integer delay     is  estimated based on the combined channel vector   $\sum_{m=1}^{3} \vert{\widehat{\mathbf h}}_{\text{eff},i}^{(m)}[n]\vert$.  Then, the channel is reconstructed  for each pilot  using    [(41), \cite{MIMOAFDMchannest}].  Denote the $ \widehat{{\mathbf H}}_{\text{eff},i}^{(m)} $ by the reconstructed channel based on the $m$th pilot, the final reconstructed channel is given by $ \widehat{{\mathbf H}}_{\text{est},i} = \frac{1}{3}\sum_{m=1}^{3} \widehat{{\mathbf H}}_{\text{eff},i}^{(m)} $.

For the proposed BEM-assisted channel estimation for ISAC symbols, the computation of  $\mathbf{R}_{\text{I}}$ and $\mathbf{R}_{\text{c}}$ can be pre-calculated based on the maximum Doppler frequency and the number of paths. Thus,  the complexity is mainly dominated by the matrix inversion in (\ref{mmse}). Moreover, since the BEM approach reduces the number of estimated parameters from $N^2$ to $QP$ compared to traditional MMSE-based methods, the complexity of the proposed scheme can be approximated as $\mathcal{O}( \vert \mathcal{N}_{\text{P}}\vert^2)$.
For the modified DR-ECE, complexity is primarily determined by (\ref{drece}) and the channel reconstruction step, approximated as $\mathcal{O}(3PN+Q)$.  Finally, the proposed KF-BEM method introduces an additional complexity of approximately $\mathcal{O}(N^2)$ due to the refined channel estimation in (\ref{e_est_}).


\begin{table}  
\footnotesize
     \caption{Simulation Parameters}
    \centering
    \begin{tabular}{c|c}
    \hline
     \hline
       \textbf{ System Parameters}  &  \textbf{ Values}  \\
        \hline
        \hline
        Center frequency & $4$ GHz \\
            \hline
       Num. of  subcarriers & $256$\\
          \hline
                 CPP length   &  $24$ \\
                 \hline
                 Subcarrier spacing  &  $ \Delta f = 15$  kHz\\
                   \hline
                      Frame length  &  $64$  symbols\\
                      
 \hline
                AFDM parameters     &  \makecell{$c_2=c_1 = \frac{K}{2N}$, \\ $K \in\{1,2,\ldots,5\}. $}\\
 \hline
  \hline
   \textbf{ Sensing Parameters}  &  \textbf{Values}  \\
   \hline
      Sensing   guard length   &  $N_{\text{G}}^{\text{S}} = 28$ \\
           \hline          
   Num. of Targets & $L = 2$ \\
     \hline
     Number. of data symbols &  $\boldsymbol{ \eta} = [0,1,2,3,4]$ \\
     \hline
   Target speed & \makecell{ $\mathcal U[v_{\text{min}},  v_{\text{max}}], v_{\text{{max}}}=$    \\   $ \{925, 462, 308,231, 185 \}$ km/h \\corresponds to $\boldsymbol{ \eta}$, and $v_{\text{{min}}} = 10.$} \\
    \hline
                    Target range  &    $\mathcal U[R_{\text{{min}}},  R_{\text{{max}}}] = \mathcal U[200, 500] $ m \\
 \hline
 LPF &  \makecell{Butterworth LPF of $20$ order \\
 } 
 \\
   \hline
    ADC sampling rate &   $f_{\text{IF}}= 4\frac{\alpha R_{\text{max}}}{\nu}   $. 
 \\
    \hline
               Sensing channel model     &  LoS with pass loss factor of $2$\\
 \hline
                Parameter estimation     &  ESPRINT \cite{ESPRIT1989} \\ 

 \hline
 \hline
   \textbf{ \makecell{Communication  \\ Parameters}}  &  \textbf{Values}  \\
   \hline
               User speed   &   $v_{\text{ue}} \in \{160,  600$\} km/h    \\
 \hline
Number of path &   $P \in \{4,  8$\}  \\
 \hline
BEM parameters &    $R=2, Q=4$  \\
 \hline
Pilot   length  & $ N_{\text{G}}^{\text{D}}=(Q+1)(l_{\max}+1)$\\
 \hline
 Doppler model &   Jakes' model \\
 \hline
 \hline
    \end{tabular}
    \label{sim_para}
 \end{table}

\section{Numerical Results}

In this section, we evaluate the sensing and communication performance of the proposed AFDM-ISAC systems.     The detailed simulation parameters are summarized in Table~\ref{sim_para}.        The parameters $N_{\text{G}}^{\text{S}}=28$, the ADC sampling rate, and the LPF parameters are chosen to satisfy the maximum detection range requirement $R_{\max}$, as specified in (\ref{Rmax}).   The proposed scheme is applicable to multiple target detection, as each target produces a distinct IF component that can be resolved provided the targets are  separated in range or velocity. Without loss of generality, $L = 2$ sensing targets are considered for illustration purposes. 

Conventional schemes require a wideband ADC operating at the full system bandwidth $f_{\text{ADC}} \geq B = N\Delta f$. In contrast, the proposed scheme only requires $f_{\text{ADC}} \geq 4\alpha \Delta R_{\max}/\nu$. With the parameters in Table I ($c_1 = \frac{3}{2N}$, $\Delta R_{\max} = 300$ m), this yields $f_{\text{ADC,min}} \approx 826.3$ kHz, which is approximately $4.6\times$ lower than $B = 3.84$ MHz.  Also,
full-duplex SIC is well recognized as one of the most challenging and costly components in monostatic ISAC systems. The proposed analog-domain sensing receiver completely eliminates this requirement through the LPF-based leakage suppression.
 \begin{figure}[t]
    \centering
\includegraphics[width=0.82\linewidth]{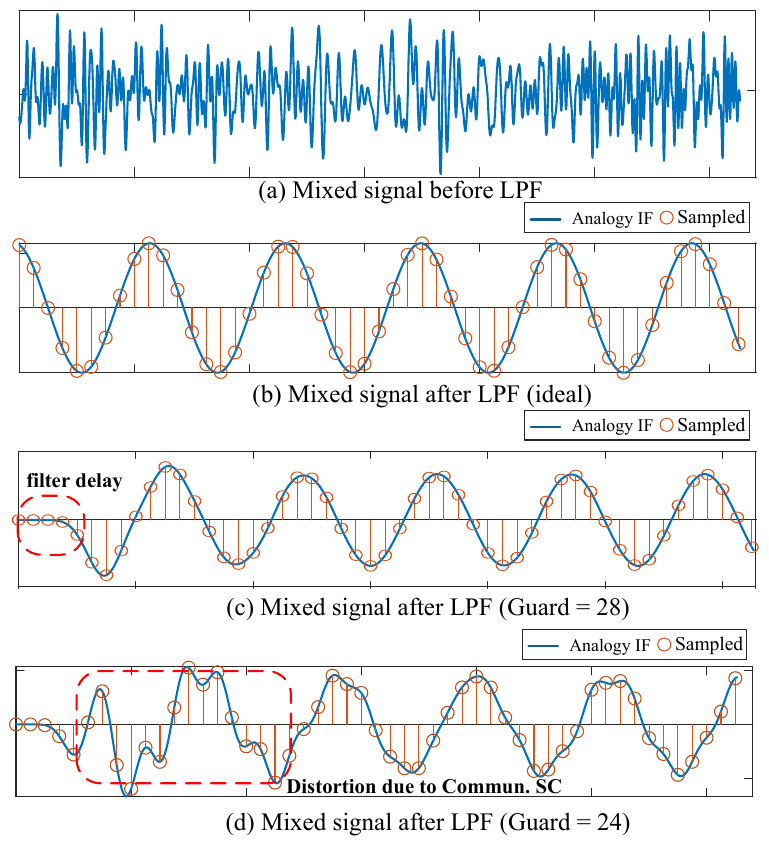}
    \caption{The dechirped signal after and before the LPF.}
    \label{LFsigna}
\end{figure}

 Fig. \ref{LFsigna}   presents an example of the original mixed signal, along with its counterparts after passing through the LPF and ADC. As  seen from   \ref{LFsigna}(b),  the proposed LPF can well eliminate the interference from the communication subcarriers. As mentioned above, the sampling frequency required by the ADC is much smaller than that of the system bandwidth. 

\subsection{Sensing Performance Comparison}

To evaluate the  sensing performance, we consider two targets whose distances are randomly drawn from a uniform distribution $\mathcal U[R_{\text{min}}, R_{\text{max}}]$. 
{We assume an  LoS  channel between the sensing targets and the ISAC platform. Accordingly, the average received SNR is modeled as     
\begin{equation}
    \text{[SNR]}_{\text{dB}} =   10 \log_{10} \left(\frac{1}{L}  \sum \nolimits_{l=0}^{L-1}    P_{\text{SPS}} R_l^{-\alpha_{\text{LoS}}} \varsigma_{\text{RCS}}  \right )/\sigma_s,     
\end{equation}
where $\varsigma_{\text{RCS}}$ denotes the   RCS attenuation  factor. The existing OFDM-, OTFS-, and AFDM-based ISAC schemes are considered as benchmark sensing schemes\cite{BemaniAFDMisac,10445333}. For a fair comparison, the number of subcarriers employed for OFDM-ISAC sensing is set to be equal to the number of guard subcarriers in the proposed AFDM-ISAC scheme. For both the conventional OTFS- and AFDM-based ISAC schemes, a dedicated sensing subcarrier with guard length $N_{\text{G}}^{\text{S}}$ is employed. Sensing is conducted based on the received pilot sections in the frequency, delay-Doppler, and affine domains for the OFDM-, OTFS-, and AFDM-based ISAC schemes, respectively. All waveforms in the simulations are generated using a raised-cosine filter with a roll-off factor of 0.1 to obtain continuous-time waveforms.

\color{black}

\begin{figure} 
	\centering
	\begin{subfigure}{0.48 \textwidth}
\includegraphics[width=0.99\linewidth]{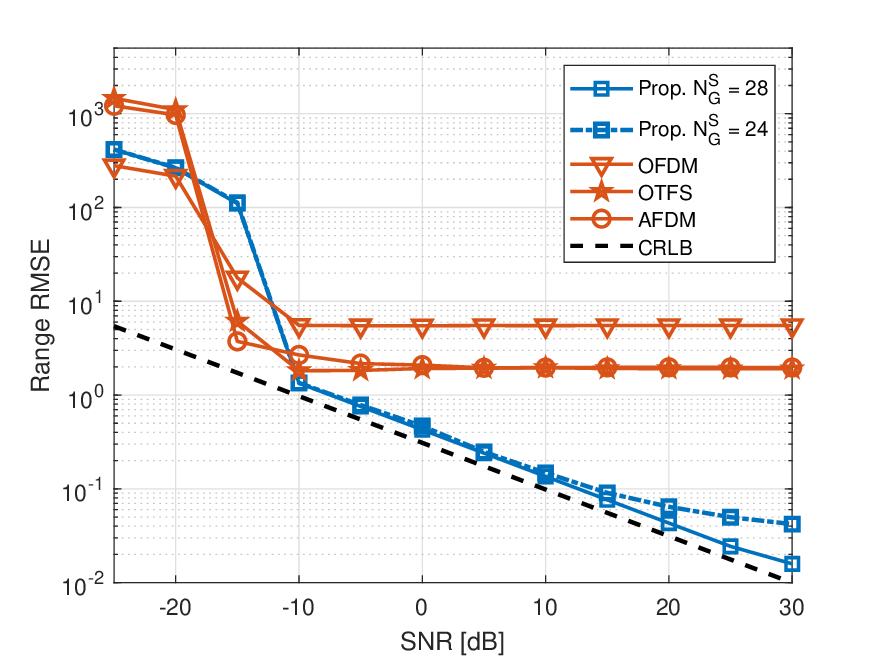}
    \caption{Range RMSE.}
				
	\end{subfigure}
	\begin{subfigure}{0.48\textwidth}
\includegraphics[width=0.99\linewidth]{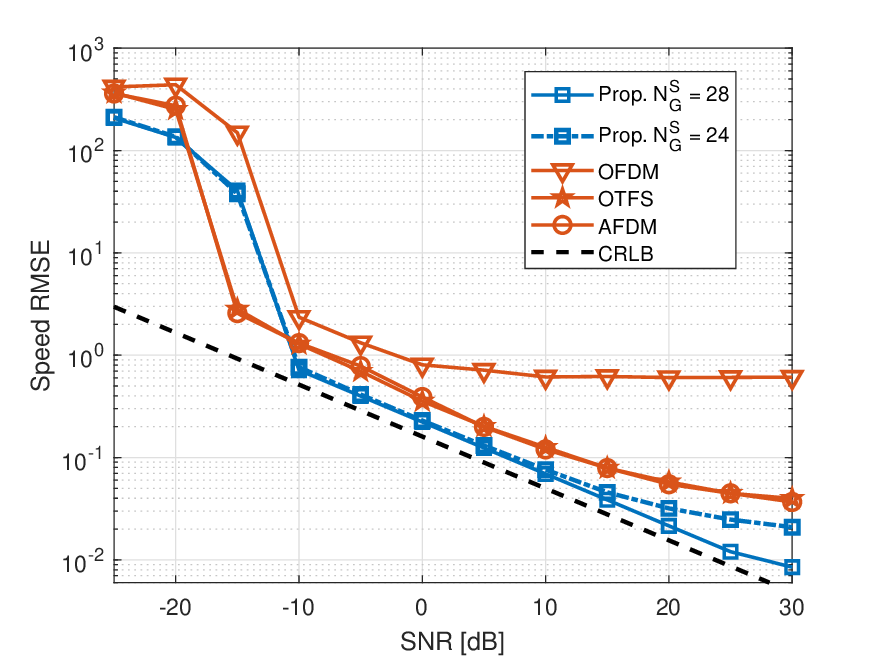}
    \caption{Speed RMSE.}
	\end{subfigure}
                 \caption{Range and speed RMSE of the proposed AFDM-ISAC ($c_1 = \frac{2}{2N}$).}
                 	\label{RSRMSE}
                      \vspace{-0.5em}    
\end{figure}


Fig.   \ref{RSRMSE} illustrates the RMSE  performance of range and speed estimations with $c_1 = \frac{2}{2N}$, based on the simulation parameters provided in Table\ref{sim_para}. For the benchmark AFDM scheme,  the same $c_1$ is adopted. From Fig. \ref{RSRMSE},  it is  observed that the proposed scheme achieves the best RMSE performance for $\text{SNR} > 10~\text{dB}$. For the proposed scheme with $N——{\text{G}}^{\text{S}} = 24$, an error floor is observed, primarily due to interference from communication subcarriers caused by Doppler spread. This effect is also evident in the IF  signal shown in Fig. \ref{LFsigna}.
In contrast, OFDM-, OTFS-, and AFDM-based benchmarks exhibit prominent error floors in RMSE, primarily caused by waveform distortion introduced during practical digital implementations, which   may not be mitigated by simply increasing the SNR.   The  proposed AFDM-ISAC scheme  operates in the analog domain  and the waveform distortion caused by pulse shaping manifests predominantly as high-frequency components in the  IF  signal. These high-frequency distortion components can be effectively suppressed by the LPF prior to ADC sampling.   Consequently, with a sufficiently large guard interval, the RMSE performance of the proposed AFDM-ISAC   approaches the CRLB. Similar trends are observed for velocity estimation RMSE.

\begin{figure}
    \centering
\includegraphics[width=0.88\linewidth]{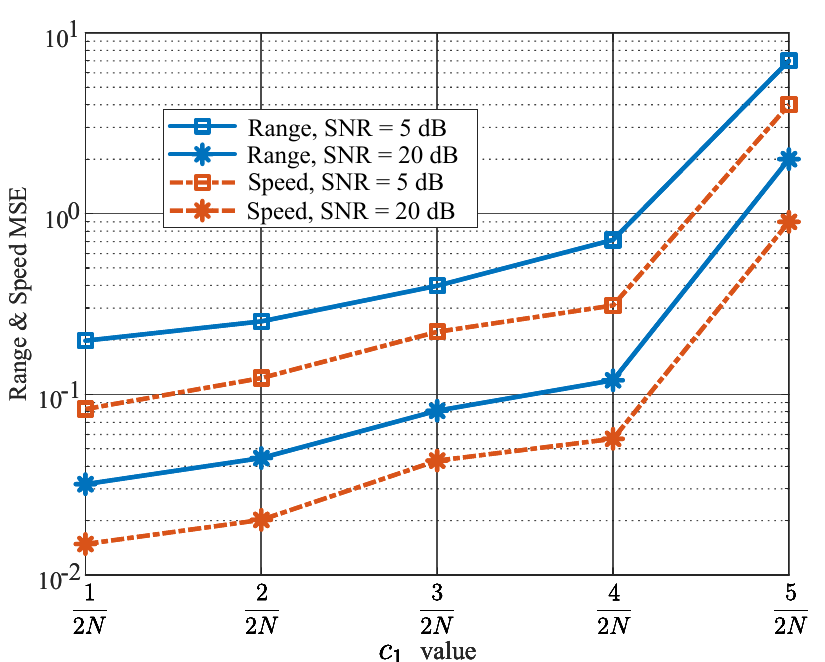}
    \caption{Range and speed RMSE for different $c_1$ values.}
    \label{c1rmse}
    \vspace{-1.5em}
\end{figure}

\begin{figure}
    \centering
\includegraphics[width=0.90\linewidth]{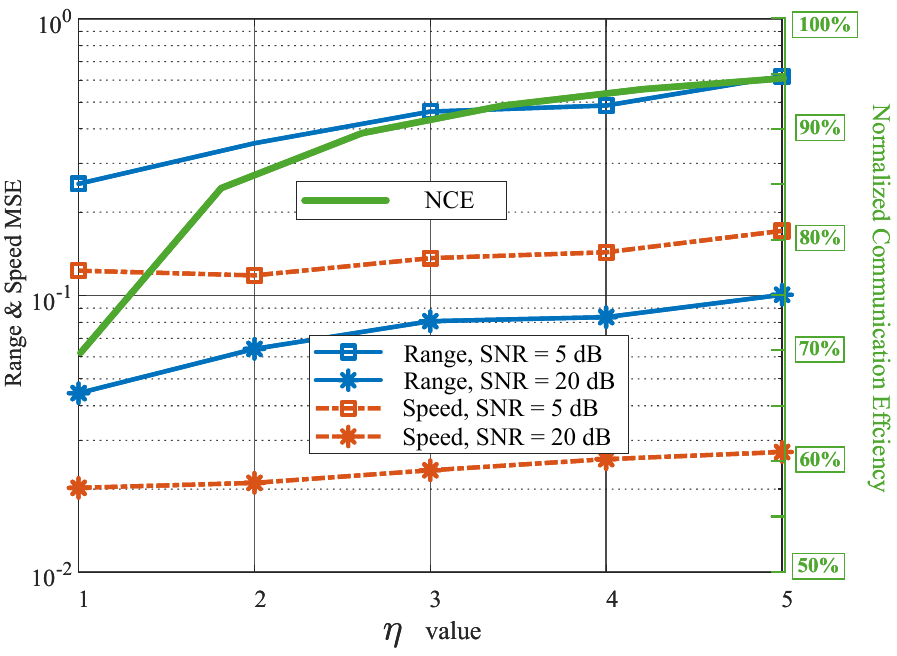}
    \caption{Trade-off between sensing performance and communication NSE via $\eta$.}
    \label{eta}
\end{figure}

  Fig. \ref{c1rmse}  further evaluates  the range and speed RMSE for  different values of  $c_1$.       It can be observed that both the range and speed RMSE increase with larger values of 
$c_1$. This degradation is primarily due to the reduced observation duration of the IF for larger $c_1$, due to  the increased propagation delay of the echoes from the sensing targets. From the perspective of sensing performance and pilot overhead, a smaller value of $c_1$ is preferred. In the next section, we will demonstrate that moderate $c_1$ values, which yield favorable sensing performance, can also achieve a desirable trade-off between sensing accuracy and communication performance.

    Fig.~\ref{eta} shows the trade-off between sensing performance and NCE, i.e., $ \eta_{\text{NCE} }$ defined in (\ref{NCE}), for different values of $\eta$. It is observed that both the range and speed RMSE degrade as $\eta$ increases. This is expected, as $\eta$ directly controls both the number of ISAC symbols and the   number of communication subcarriers. The proposed AFDM-ISAC scheme features a flexible frame structure that enables NCE enhancement by appropriately tuning $\eta$. Since ISAC symbols also carry data subcarriers, the scheme maintains favorable NCE performance even when $\eta = 0$, i.e., when all symbols are configured as ISAC symbols. More importantly, increasing $\eta$ leads to only a slight degradation in speed RMSE while achieving a significant gain in NCE. This trade-off is reasonable because increasing $\eta$ effectively downsamples the IF signal in the Doppler domain. Provided that  the maximum target speed satisfies $v_l \leq v_{\text{max}}$, the speed RMSE remains within acceptable bounds.

\begin{figure}
    \centering
\includegraphics[width=0.88\linewidth]{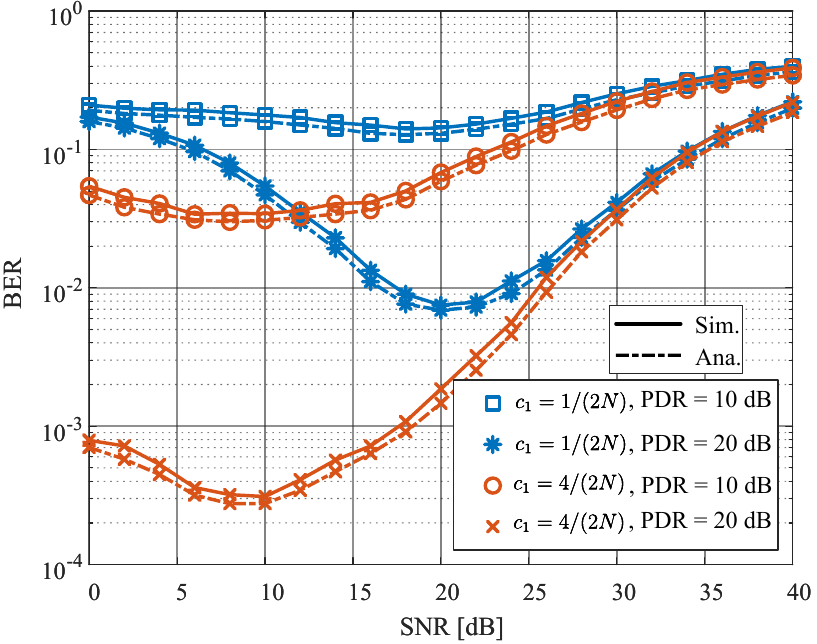}
    \caption{Simulated  and analytic BER performance for different power allocation factors.}
    \label{powerallo}
\end{figure}

  \begin{figure*}[]
	\centering
	\begin{subfigure}{0.48 \textwidth}
\includegraphics[width=0.88\linewidth]{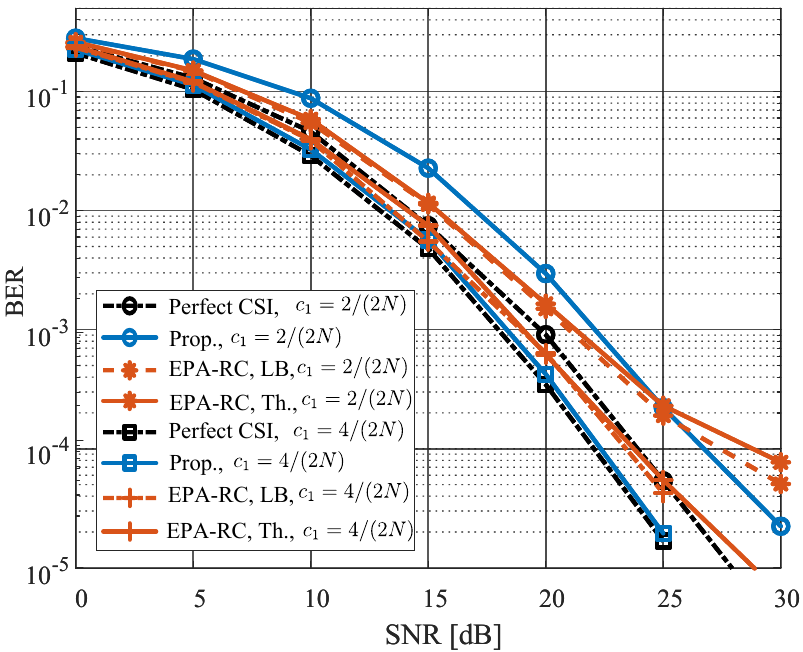}
    \caption{$160$ km/h.}
	\vspace{-0.1em}
	\end{subfigure}
	\begin{subfigure}{0.48\textwidth}
\includegraphics[width=0.9\linewidth]{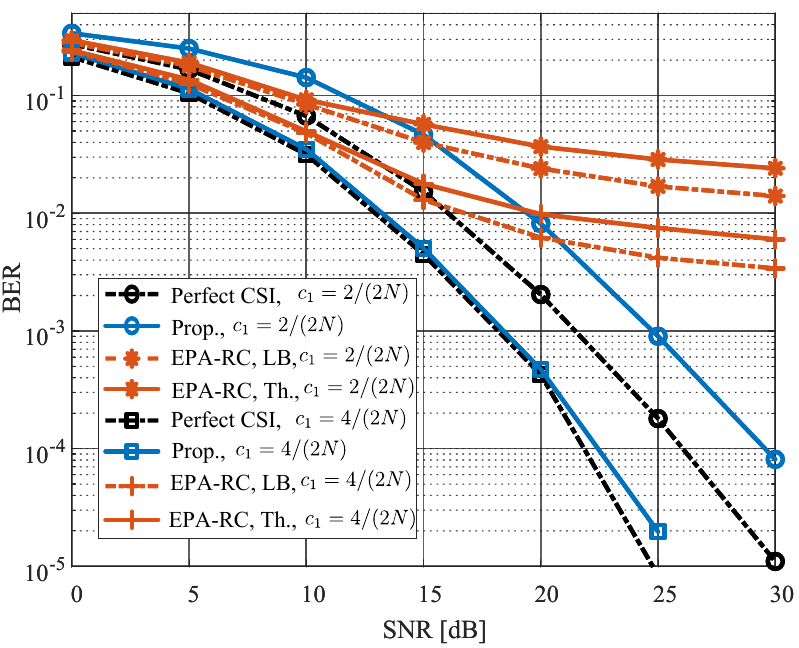}
                 \caption{$600$ km/h.}
		\vspace{-0.2em}
	\end{subfigure}
                 \caption{BER comparisons of different channel estimation schemes.}
                 	\label{BER_CHANEL}
                    \vspace{-2em}
\end{figure*}

\subsection{Communication Performance}
In this section, we evaluate the communication performance of the proposed AFDM-ISAC system. The detailed simulation parameters are summarized in Table~\ref{sim_para}. For simplicity, we set $Q = 4$ for both low- and high-mobility user scenarios. Unless otherwise stated, $\eta = 0$ is assumed in Figs.~\ref{powerallo}–\ref{BER_COMM2}.

Fig.~\ref{powerallo} presents both the simulated and analytical BER performance under varying pilot-to-data ratio  (PDR), defined as $\small \text{PDR}= 10\log \frac{P_{\text{SP}}}{P_{\text{d}} } \quad \text{[dB]} $,    at a user mobility of $600$ km/h. As observed, the BER initially improves with increasing $\text{SNR}$, owing to more accurate channel estimation. However, beyond the optimal point, the BER begins to degrade as less power is allocated to the data symbols. Furthermore, when the modulation parameter $c_1$ takes a larger value (e.g., $c_1 = \frac{4}{2N}$), the optimal BER can be achieved with relatively lower pilot power, indicating more efficient power utilization.

\begin{figure}
    \centering
\includegraphics[width=0.92\linewidth]{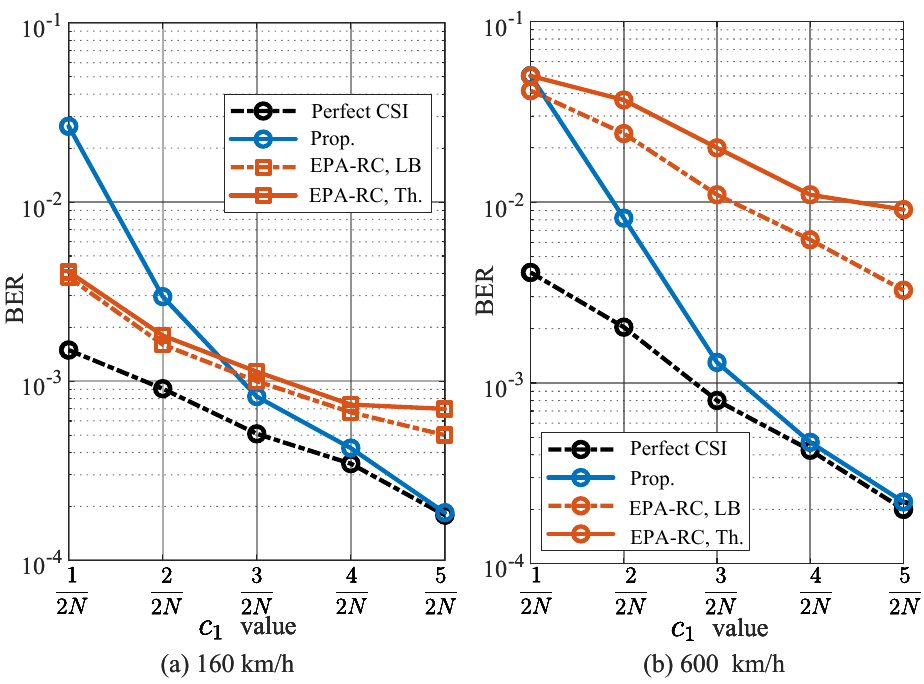}
    \caption{BER v.s $c_1$ values at SNR$=20$ dB.}
    \label{BER_COMM2}
      \vspace{-0.5em}
\end{figure}

Fig.~\ref{BER_CHANEL} compares the BER performance of the proposed BEM-assisted scheme with the EPA-RC channel estimation method under user mobilities of $160$ km/h and $600$ km/h. In this comparison, ``EPA-RC, Th" refers to  a threshold-based method is used to estimate the path locations, as described in (\ref{drece}), while ``EPA-RC, LB" represents a performance lower bound, assuming perfect knowledge of the path locations at the receiver. The key observations from the results are summarized as follows:
\begin{itemize}
    \item The gap between the ``EPA-RC, LB” and ``EPA-RC, Th” schemes is small, indicating that the threshold in this paper is well designed. While EPA-RC slightly outperforms the proposed BEM-assisted channel estimation method for $c_1 = \frac{2}{2N}$ at $160$ km/h in the low SNR regime, the proposed scheme demonstrates a clear advantage at higher mobility. In particular, under a user mobility of $600$ km/h, the proposed approach significantly outperforms the EPA-RC method across the entire SNR range.
    \item An SNR gap between $c_1 = \frac{2}{2N}$ and $c_1 = \frac{4}{2N}$ is observed under the perfect channel state information (CSI) case. This is expected, as more power is allocated to pilot symbols in the $c_1 = \frac{2}{2N}$ setting, which reduces the effective SNR available for data symbols and consequently degrades BER performance.
    \item Our proposed BEM-assisted channel estimation approach achieves robust BER performance even at a user mobility of $600$ km/h. Notably, when $c_1 = \frac{4}{2N}$, the proposed scheme closely approaches the performance of the perfect CSI benchmark, highlighting its effectiveness in high-mobility scenarios.

\end{itemize}

\begin{figure}
    \centering
\includegraphics[width=0.86\linewidth]{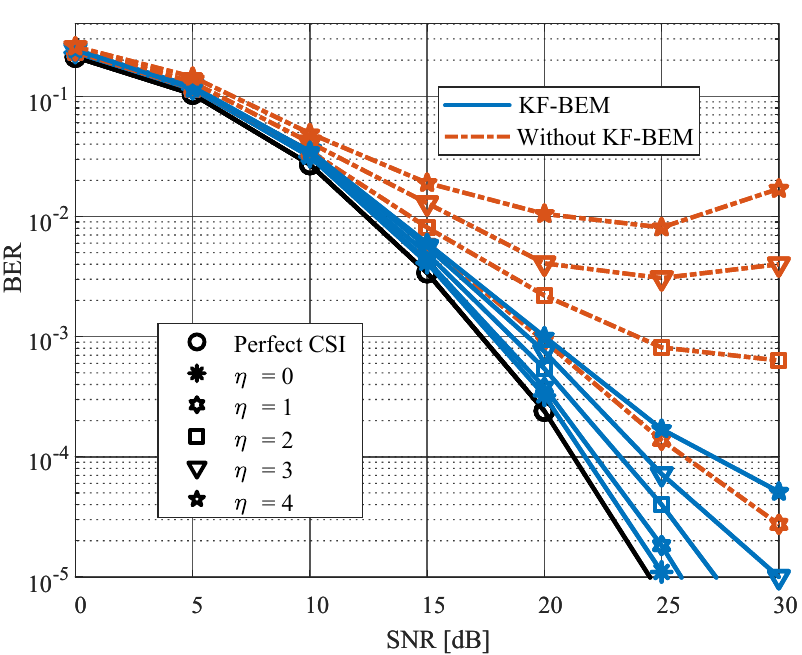}
    \caption{BER performance of the proposed KF-BEM for different $\eta$.}
    \label{EFBEM}
 \end{figure}

Fig.~\ref{BER_COMM2} further illustrates the BER performance across different values of $c_1$ at $\text{SNR} = 20$ dB. One can see that the proposed BEM-assisted channel estimation progressively approaches the performance of the perfect CSI case as $c_1$ increases. In particular, the performance gap becomes negligible for $c_1 \geq \frac{4}{2N}$. Additionally, the BER under perfect CSI also improves   with increasing $c_1$. The reasons are two-fold: 1) larger $c_1$ values require less pilot power for reliable channel estimation, allowing more transmit power to be allocated to data symbols; and 2) the AFDM waveform inherently offers better BER performance at higher $c_1$, even when the data symbol power is fixed. 
Furthermore, while the EPA-RC channel estimation slightly outperforms the proposed scheme for $c_1 = \frac{1}{2N}$ and $\frac{2}{2N}$ at a user mobility of $ 160$ km/h, its performance degrades significantly in high-mobility scenarios (e.g., $600$ km/h). Based on Fig.~\ref{c1rmse}, $c_1 \in \left\{ \frac{2}{2N}, \frac{3}{2N}, \frac{4}{2N} \right\}$ yields a  {desirable trade-off between sensing accuracy and communication performance under the considered system configurations.}

 Fig.~\ref{EFBEM} shows the BER performance of the proposed KF-BEM channel estimation scheme, where $c_1 = \frac{4}{2N}$ and a user speed of $600$ km/h are considered. As can be seen from the figure, the proposed KF-BEM scheme maintains favorable BER performance for $0 \leq \eta \leq 3$. However, for $\eta = 4$, a noticeable performance degradation appears at high SNRs due to increased channel variation.  More importantly, in the absence of KF-BEM tracking, the BER performance degrades significantly as $\eta$ increases, and an error floor is observed at high SNRs, primarily due to error propagation caused by inaccurate channel estimation. These results show the theoretical advantage of the proposed KF-BEM: since the Kalman filter recursively fuses the predicted state from (\ref{eqc}) with the current observation via (\ref{eqsu}), it yields a lower or equal  MSE estimation compared to the one-shot LMMSE estimator in (\ref{mmse}). This gain becomes more pronounced as $\eta$ increases, since pure data symbols lack dedicated pilots and must rely entirely on temporal prediction and tracking.

 \begin{figure}
    \centering
\includegraphics[width=0.95\linewidth]{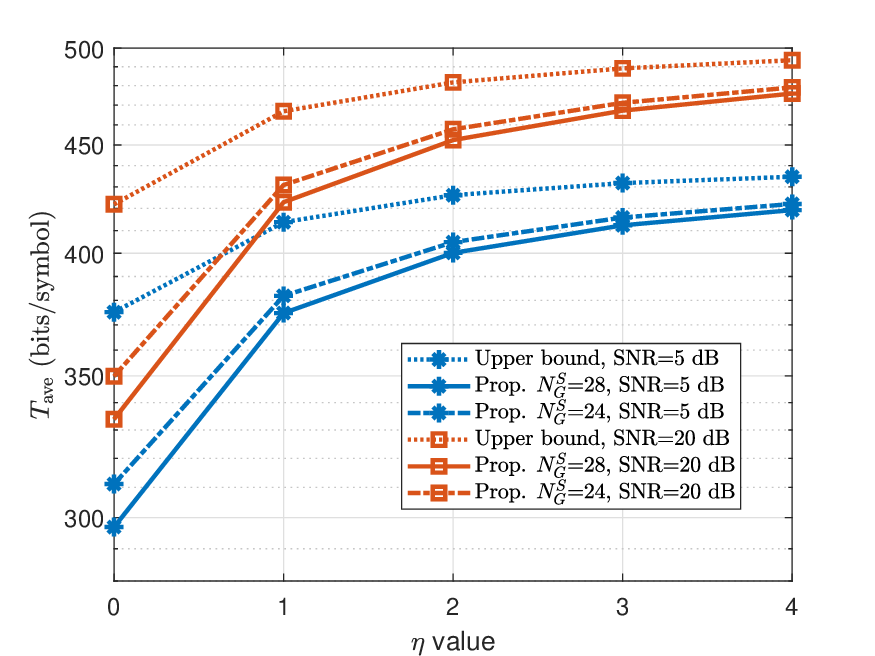}
    \caption{ Average throughput ($  T_{\text{ave} }$) comparison for different $\eta$ and $N_{\text{G}}^{\text{S}}$ values.}
    \label{AT}
    \vspace{-0.5em}
 \end{figure}

Finally, Fig.~\ref{AT} compares the achieved average throughput of the proposed AFDM-ISAC frame structure with the KF-BEM, which is defined as 
\begin{equation}
    \small
    \label{TR}
    \begin{aligned}
           T_{\text{ave} }&=   \log_2(M) (1-\text{BER}) \underbrace{\frac{N-2N_{\text{G}}^{\text{D}}-2N_{\text{G}}^{\text{S}}-3+ \eta N}{1+\eta}}_{\text{Num. of Commun. Subcarriers}}.
    \end{aligned}
\end{equation}
The curves labelled ``upper bound'' in Fig.~\ref{AT} correspond to the case where all subcarriers are allocated for communication, i.e., $N_{\text{G}}^{\text{S}} = 0$.   It can be observed from Fig.~\ref{AT} that a throughput loss of approximately $50$ bits/symbol is incurred when $\eta = 0$, which corresponds to the configuration that achieves the best sensing performance, as shown in Fig.~\ref{eta}. However, as $\eta$ increases, the throughput of the proposed scheme progressively approaches the upper bound, provided that the maximum target speed remains within the unambiguous range defined in (\ref{vmax}). This further shows the trade-off of  the proposed AFDM-ISAC frame structure in between sensing accuracy and communication throughput through the tuning of $\eta$ and $N_{\text{G}}^{\text{S}}$.

\color{black}


\section{Conclusion}

 In this paper, we   proposed a practical and flexible  AFDM-ISAC  framework.  A novel frame structure consists of ISAC symbol and pure data symbol was first designed, upon which  each ISAC symbol embeds a single chirp subcarrier for both   sensing and channel
estimation. This enables full-bandwidth sensing with minimum sacrificing of communication rate. To overcome hardware limitations and reduce system complexity, an analog-domain sensing receiver equipped with a carefully designed LPF was also  introduced. Our proposed  design not only suppresses interference from communication subcarriers but also mitigates ADC saturation, thus eliminating the need for costly full-duplex circuits. Furthermore, an AFDM parameters-guided sensing aggregation algorithm    was proposed   in the digital domain to enhance sensing robustness. On the communication side, we   proposed a GCE-BEM-assisted low complexity channel estimation scheme. Power allocation between pilot and data symbols was optimized, and a GCE-BEM-based Kalman filter was developed for reliable frame-based channel estimation. Extensive simulation results validated the proposed AFDM-ISAC design, demonstrating superior performance over existing OFDM, OTFS, and AFDM baselines in terms of sensing accuracy, communication reliability, hardware simplicity, and system flexibility.  In particular, it was found that $c_1 \in  \{ \frac{2}{2N},\frac{3}{2N},\frac{4}{2N}\}$ and choosing the pure-data-to-ISAC symbol ratio  $\eta \in \{1,2,3 \}$ can achieve favorable performance in both sensing and communication.

Future work includes extending the proposed framework to 
multi-antenna  systems for three-dimensional radar estimation 
(e.g., angle of arrival/departure), generalizing to bistatic 
and multistatic configurations, investigating nonlinear impairments such as Doppler scaling 
errors and hardware nonlinearities, and validating the 
proposed design through hardware prototyping.

 \vspace{-1em}

\ifCLASSOPTIONcaptionsoff
  \newpage
\fi
\appendices

 \section{Derivation of $\mathbf R_{\text{c}}$ and  $\mathbf R_{\text{d}}$}  
\label{eqRcd}
\subsubsection{Derivation of $\mathbf R_{\text{c}}$}  Let us rewrite (\ref{h_ti})  in the vector form as $\mathbf h_{i,p} = \mathbf B \mathbf c_{i,p}$, where $\mathbf B = [ \mathbf b_0, \mathbf b_1, \ldots, \mathbf b_{N-1}]^{\mathcal T}$. Then, one has  $\mathbf c_{i,p} = \left (\mathbf B^{\mathcal H} \mathbf B \right)^{-1} \mathbf B^{\mathcal H}\mathbf h_{i,p}.$ Accordingly,  the corelation matrix  of the BEM coefficient at the $p$th can be expressed as   
\begin{equation}
\small
    \mathbf R_{\text{c},i,p} \! = \! \mathbb E\{ \mathbf c_{i,p}\mathbf  c_{i,p}^{\mathcal H} \} =  \Big (\mathbf B^{\mathcal H} \mathbf B \Big)^{-1} \mathbf B^{\mathcal H}  \mathbf R_{\text{h},i,p}\mathbf B  \Big(\mathbf B^{\mathcal H} \mathbf B \Big)^{-1},
\end{equation}
where $\mathbf R_{\text{h},i,p}=  \mathbb E\{ \mathbf h_{i,p}\mathbf  h_{i,p}^{\mathcal H} \} $.  In this paper, we consider channel taps as complex Gaussian processes of variances of zero mean and with wide sense $\sigma_{\text{h},p}$, which follow the Jakes power spectrum of the maximum Doppler frequency $ f_{\text{d, max}}$, then
\begin{equation}
\small
\begin{aligned}
        \left[\mathbf R_{\text{h},i,p}\right]_{n,m} & =  \mathbb E\{  h_{i}(m,p)  h_{i}^{*}(m,l) \} \\ & =  \sigma_{\text{h},p}J_{0}\left (   2\pi  f_{\text{d,max}}(m\!-\!n)T_{\text{s}} \right  ),  
\end{aligned} 
\end{equation}
where $ J_{0}(\cdot)$ is the Bessel function of first kind and order zero. 
  
\subsubsection{Derivation of $\mathbf R_{\text{d}}$}       Denote $\mathbf D_1 =  [  \mathbf A   \mathbf B^0 \textbf {F}^{\mathcal H}, \mathbf A   \mathbf B^1 \textbf {F}^{\mathcal H}, \ldots, \mathbf A    \mathbf B^{Q-1} \textbf {F}^{\mathcal H} ]  $,  $\mathbf D_2 =  \mathbf I_{Q} \otimes \text {diag}\{\textbf {F} \mathbf A^{\mathcal H}_{\text{d}}  \mathbf {x}_{i,\text{d}}\} \mathbf {F}_{P} $. Then, one has 
\begin{equation}
\small
\begin{aligned}
      \mathbf R_\text{d} & = \mathbb E \{\widetilde{ \mathbf M}_{i, \text{d}} \textbf {c}_i    \mathbf {c}_i^{\mathcal H} (\widetilde{ \mathbf M}_{i, \text{d}}  )^{\mathcal H}\}   =    \mathbf D_1  \mathbb E \{ \mathbf D_2    \mathbf {c}  \mathbf {c}^{\mathcal H}  \mathbf  D_2 ^{\mathcal H}\}  \mathbf D_1^{\mathcal H}  \\ 
      & = \mathbf D_1  \mathbb E \{  \mathbf I_{Q+1}\otimes\text{diag}  \{\textbf {F} \mathbf A^{\mathcal H} \mathbf {x}_{i,\text{d}} \}  \mathbf I_{Q}  \otimes \mathbf F_L  \mathbf {c}_i  \mathbf {c}_i^{\mathcal H}   \\
      & \quad  (\mathbf I_{Q}  \otimes \mathbf F_P )^{\mathcal H} ( \mathbf I_{Q}\otimes\text{diag}  \{\textbf {F} \mathbf A^{\mathcal H}   \mathbf {x}_{i,\text{d}} \})^{\mathcal H}  \} \mathbf D_1^{\mathcal H} \\
          & = \mathbf D_1  (\mathbf I_{Q}\otimes \textbf {F} \mathbf A^{\mathcal H}_{\text{d}} \sigma_{\text{d}}  \mathbf A_{\text{d}} \mathbf F^{\mathcal H})  \odot \mathbf D_3    \mathbf D_1^{\mathcal H},\\
\end{aligned}
\end{equation}
where $ \mathbf A_{\text{d}}  $ denotes the columns of $\mathbf A$ associated with the data subcarriers ${\text{Ind}_{\text{d}}}$, and   $ \mathbf D_3 =  \mathbf I_{Q}  \otimes \mathbf F_P   \mathbf R_{\text{c},i} ( \mathbf I_{Q}  \otimes \mathbf F_P  )^{\mathcal H}  $.    
 


%

\bibliography{ref.bib} 
\bibliographystyle{IEEEtran}

\end{document}